\documentclass[preprint2]{aastex62}

\usepackage{graphicx}
\usepackage{subfigure}
\usepackage{longtable}
\usepackage{wrapfig}
\usepackage[figuresleft]{rotating}\usepackage{epstopdf}
\usepackage{amsmath}
\usepackage{multirow}
\usepackage[mathscr]{eucal}

\newcommand{\kms}{\mbox{${\rm km~s^{-1}}$} }

\newcommand{\hi}{H{\sc\,i} }
\newcommand{\dg}{$^{\circ}$ }
\newcommand{\Lc}{\mathcal{L}}
\newcommand{\Pc}{\mathcal{P}}
\newcommand{\Dc}{\mathcal{D}}
\newcommand{\Mc}{\mathcal{M}}

\graphicspath{{./}{figures/}}

\submitjournal{ApJ}

\shorttitle{Global Attenuation in Spiral Galaxies}
\shortauthors{Kourkchi et al.}

\begin{document}

\title{Global Attenuation in Spiral Galaxies in Optical and Infrared Bands}

%%\correspondingauthor{Ehsan Kourkchi}
\email{ehsan@ifa.hawaii.edu}

\author{Ehsan Kourkchi}
\affil{Institute for Astronomy, University of Hawaii, 2680 Woodlawn Drive, Honolulu, HI 96822, USA}
\author{R. Brent Tully}
\email{tully@ifa.hawaii.edu}
\affil{Institute for Astronomy, University of Hawaii, 2680 Woodlawn Drive, Honolulu, HI 96822, USA}
\author{J. Don Neill}
\affil{California Institute of Technology, 1200 East California Boulevard, MC 278-17, Pasadena, CA 91125, USA}
\author{Mark Seibert}
\affil{The Observatories of the Carnegie Institute of Washington, 813 Santa Barbara Street, Pasadena, CA 91101, USA}
\author{H\'el\`ene M. Courtois}
\affil{University of Lyon, UCB Lyon 1, CNRS/IN2P3, IPN Lyon, France}
\author{Alexandra Dupuy}
\affil{University of Lyon, UCB Lyon 1, CNRS/IN2P3, IPN Lyon, France}

\begin{abstract}

The emerging light from a galaxy is under the influence of its own interstellar medium, as well as its spatial orientation. Considering a sample of 2,239 local spiral galaxies in optical ({\it SDSS $u$, $g$, $r$, $i$, and $z$}) and infrared bands ({\it WISE $W1$, $W1$}), we study the dependency of the global intrinsic attenuation in spiral galaxies on their morphologies, sizes and spatial inclinations.
Reddening is minimal at the extremes of low mass and gas depletion and maximal in galaxies that are relatively massive and metal rich and still retain substantial gas reserves.  A principal component constructed from observables that monitor galaxy mass, relative \hi content to old stars, and infrared surface brightness is strongly correlated with the amplitude of obscuration. We determine both a parametric model for dust obscuration and a non-parametric model based on the Gaussian process formalism. An average dust attenuation curve is derived for wavelengths between 0.36 and 4.5~$\mu$m.

\end{abstract}



\section{Introduction} \label{sec:intro}

A large new data set provides insights into the properties of obscuration in spiral galaxies.  The data have been accumulated toward the goal of determining galaxy distances through the correlation between the intrinsic luminosities and rotation rates of spiral galaxies \citep{1977A&A....54..661T}.  It is important to understand the consequences of obscuration on observed luminosities for that purpose.  In addition to the utilitarian use for the measurement of distances, there is a general interest in understanding the statistical properties of obscuration in spiral galaxies.

The light from stars is attenuated\footnote{Following advice from the referee, we reserve the term ``extinction" to refer to the local interplay of absorption of scattering of light with the interstellar medium and use the term ``attenuation" for the global effect dependent on source geometry and attendant path lengths.} by intervening dust in the line of sight through a host galaxy.  Dust accumulation is enhanced in metal-rich environments where the interstellar medium has been retained.  Attenuation of optical light is insignificant at the extremes of dwarf galaxies and galaxies overwhelmingly dominated by old stars.  Dwarf galaxies can have considerable gas content but most is in the atomic state, transparent to visible light.  The gas in predominantly old systems has been efficiently transformed into stars or purged through winds or stripping.  In between in terms of mass and evolution, there are normal spiral galaxies with orientations that can hide two-thirds of their stellar light at blueward passbands.  {\it It will be demonstrated that the degree of obscuration from a non-pathological galaxy is predictable, given a practical suite of observables.}

The stellar light of galaxies is obscured through interactions with their interstellar gas and dust. The level of obscuration correlates with spatial orientation and physical properties of the host galaxy including size or mass, morphology, formation history and metallicity or the chemical compositions of its interstellar environment \citep{2011piim.book.....D}. The effect of dust obscuration is less pronounced at longer wavelengths and on average it effectively shifts the galaxy color redward. The amplitude of dust obscuration in spirals increases with their inclination from face-on, because interactions between the galaxy photons and its dust/gas particles occur along longer paths in the direction of the observer line-of-sight \citep{2010MNRAS.404..792M}. There have been 
various attempts to study the influence of dust on the observed luminosity of spirals at multiple wavelengths \citep{1995AJ....110.1059G,1998AJ....115.2264T, 2003AJ....126..158M, 2008ApJ...687..976U}. \citet{2009ApJ...693.1045C} describe dust attenuation in late-type galaxies in terms of a concentration index and $Ks$-band absolute luminosity. \citet{2010ApJ...709..780Y} study the dependency of dust attenuation in spirals on their orientations from measurements of Balmer line ratio, $H\alpha / H\beta$.
\citet{2012MNRAS.421..486X} find an empirical relation between dust attenuation and such physical parameters of spirals as $H\alpha$ luminosity, $H\alpha$ surface brightness, metallicity and axial ratio.

The information we bring to the problem begins with the kinematic information derived from global 21cm neutral Hydrogen profiles. HI linewidths are a distance independent proxy for intrinsic luminosities \citep{1998AJ....115.2264T}. The HI observations also provide fluxes. Pseudo-colors formed from optical/infrared photometry and HI fluxes are distance independent and are correlated with morphology, stellar populations, interstellar gas content, and presumably the accumulation of dust. 

In addition, we have multi-band surface photometry: using the $u$,$g$,$r$,$i$,$z$ bands from the archive of the Sloan Digital Sky Survey ($SDSS$) and the $W1$,$W2$ (3.4 and 4.6 $\mu$m) bands from imaging with the Wide-field Infrared Survey Explorer ($WISE$). Colors formed by combinations of these bands are independent of distance and bear on stellar populations $-$ ages and metallicities $-$ as well as the effects of reddening.  Photometry also provides two other distance-independent parameters that reflect the morphologies of galaxies: surface brightness and degree of central concentration.

Finally, there is galaxy inclination, also a distance independent parameter. Cumulative obscuration increases sharply in galaxies toward the edge-on orientation. We have carefully measured the inclinations of our spirals by manually comparing them to a set of standard spirals with known inclinations. We believe our measured inclinations are more accurate compared to those calculated from the axial ratio of photometry iso-density ellipses which are subject to morphological peculiarities, and in some cases they may represent the galaxy bulge instead of its spiral disk.

The problem posed to us is to disentangle the matrix of these observables with colors from multi-band optical, infrared, and \hi fluxes on one axis of the matrix and the structural parameters, surface brightness, and concentration, line profile width intrinsic luminosity proxies, and inclinations on the other axis using a sample of 2,239 galaxies. We monitor the amplitude of obscuration as a function of stellar and gas properties as manifested by inclination dependencies.  Our parameters are all distance-independent which enables an analysis with a sample large enough to characterize the dominant tendencies. 

We present our data products in \S \ref{sec:data}. 
In \S \ref{sec:iog}, we introduce a set of distance independent observables, and the main principal component combination of observables that minimizes dispersion in optical$-$infrared colors, to address the problem of dust obscuration. 
In \S \ref{sec:method}, we take parametric and non-parametric approaches to formulate the inclination dependent dust attenuation. In \S \ref{sec:rvw} we calculate the average dust attenuation in spirals at different wavelengths. We summarize our results in \S \ref{sec:summary}.

\section{Data} \label{sec:data}

The sample for this study comprises 2,239 spiral galaxies taken from a grand catalog of $\sim 20,000$ spiral galaxies {\footnote This catalog would be presented in a following paper.} assembled for the measurement of distances using the correlation between galaxy \hi linewidths and luminosities known as the Tully-Fisher Relationship, hereafter TFR: \citep{1977A&A....54..661T}. 
Galaxies of this grand catalog were selected based on the following criteria: (1) All lie within 15,000 \kms, most within the declination range accessible to the Arecibo Telescope of $0 < \delta < +38$. (2) Morphological types are {\it Sa} or later. (3) Their preliminarily estimated inclinations are greater than 45\dg from face-on based on axial ratios taken from HyperLEDA\footnote{\url{http://leda.univ-lyon1.fr/}} \citep{2003A&A...412...45P}. (4) All are restricted to have high quality \hi measurements as explained in \S \ref{subsection:HI}.

The current study of dust attenuation requires spirals to have both optical and infrared photometry coverage that is not currently available for the full sample of $\sim 20,000$ galaxies. We have completed the optical photometry of all galaxies with SDSS coverage, but the infrared photometry of the full sample was not completed at the time of this study. The requirement of complete photometry in all bands (optical and infrared) left us with a sub-sample of 2,239 galaxies for use in this study.

The neutral Hydrogen linewidths and fluxes, sources and criteria used for the compilation of our sample are discussed in \S \ref{subsection:HI}. Our procedures to obtain and analyze optical and infrared images are explained in \S \ref{subsec:optical}, \ref{subsec:infrared}, and \S \ref{subsec:photometry}.  Regarding inclinations, axial ratios are not sufficiently accurate indicators of spiral inclinations so we only used such estimates to give a preliminary sample selection cut at $45^{\circ}$.  More face-on spirals are excluded because uncertainties in linewidth deprojection become prohibitive. In \S \ref{subsec:inclination} we discuss how we proceeded to acquire inclinations of reasonable quality and that supersede inclinations derived from axial ratios. In \S \ref{subsec:catalog}, we present our data catalog.

\subsection{Neutral Hydrogen} \label{subsection:HI}

\hi linewidth and flux measurements of sufficient quality are provided by the All Digital \hi catalog ({\it ADHI}) that has been assembled in the context of the $Cosmicflows$ program \citep{2016AJ....152...50T} and is available at the Extragalactic Distance Database (EDD) website\footnote{\url{http://edd.ifa.hawaii.edu}; catalog ``All Digital \hi''.} and/or the Arecibo Legacy Fast ALFA Survey ({\it ALFALFA}; \citealt{2011AJ....142..170H, 2018ApJ...861...49H}). The ADHI catalog provides $W_{mx}$, \hi linewidths that represent positive and negative extremes of galaxy rotation velocities, a parameter derived from the directly observed $W_{m50}$, the linewidth at 50\% of the mean \hi flux within the area that contains 90\% of the total \hi flux \citep{2011MNRAS.414.2005C, 2012ApJ...749...78T, 2012ApJ...749..174C}. We adjust {\it ALFALFA} linewidths, $W_{alf}$, for compatibility with the {\it ADHI} values using $W_{mx}=W_{alf}-6$ \kms, which we derived for galaxies covered by both catalogs. We use a weighted average of \hi measurements if a galaxy is included in both catalogs. Our interest is in spiral, not dwarf, galaxies so we set a threshold of $W_{mx}>64$ \kms. To avoid marginal data we require $S/N>10$ for linewidths taken from the {\it ALFALFA} catalog and we only accept {\it ADHI} $W_{mx}$ values with uncertainties less than or equal to $20$ \kms. These two criteria are chosen to be fairly consistent for a set of common galaxies in both catalogs. Coupled with the availability of high quality optical/infrared photometry and inclination measurements, we draw 2,004 galaxies from ADHI and 1,248 galaxies from ALFALFA, many in common. A further 61 \hi linewidth measurements were extracted from \citet{2005ApJS..160..149S} where their linewidth values $W_{M50}$ are adjusted to our $W_{m50}$ using the linear relationship $W_{m50} - W_{M50} = 1.015W_{m50}-11$ \citep{2009AJ....138.1938C}. To summarize, out of the final 2,239 spirals in our sample, 61 have \hi measurements from \citet{2005ApJS..160..149S} and the rest are covered by {\it ADHI} and/or {\it ALFALFA} catalogs.

\hi fluxes are derived from the average of all available fluxes drawing on these resources.
Analogous to the magnitude scale used to characterize optical and near infrared fluxes, the 21~cm line flux translates to  the magnitude $m_{21}$ with the relation
\begin{equation}
    m_{21} = -2.5 {\rm log} F_{HI} + 17.40,
\end{equation}
where $F_{HI}$ is the area of the 21-cm line profile in the units Jy$\cdot$\kms.

\subsection{Optical Images} \label{subsec:optical}

To obtain the cutout image of each galaxy at {\it u,g,r,i} and {\it z} bands, we download all corresponding calibrated single exposures from the {\it SDSS DR12} database \citep{2000AJ....120.1579Y}. All exposures have already been calibrated and sky-subtracted. Therefore, we do not perform any sky-subtraction prior to combining frames. We use {\tt MONTAGE}, a toolkit for assembling astronomical images \citep{2010ascl.soft10036J}, to drizzle all frames and construct galaxy images. The angular scale of the output images is 0.4'' pixel$^{-1}.$ Our data acquisition pipeline is accessible online\footnote{\url{https://github.com/ekourkchi/SDSS\_get}}. SDSS provides images that are calibrated in such a way that all measured magnitudes are supposed to be in the AB system in all {\it ugriz} bands. Although this is correct for {\it gri} bands, there are small departures for {\it SDSS} {\it u} and {\it z} band zero points from the AB system that are corrected using $u_{AB}=u_{SDSS}-0.04$ mag and $z_{AB}=z_{SDSS}+0.02$ mag\footnote{\url{https://www.sdss.org/dr12/algorithms/fluxcal/}}.

\subsection{Infrared Images} \label{subsec:infrared}

We use the {\it W1} (3.4$\mu m$) and {\it W2} (4.6$\mu m$) images of the {\it WISE} survey \citep{2010AJ....140.1868W}. All {\it WISE} images are available to the public through the NASA/IPAC infrared science archive (IRSA). Following the same procedure explained by \citet{2014ApJ...792..129N}, image products were constructed by drizzling single exposure images using version 3.8.4 of the Image Co-addition with Optional Resolution Enhancement ({\tt ICORE}) software \citep{2009ASPC..411...67M, 2013ascl.soft02010M}. All of our final co-added images have a spatial scale of 1'' pixel$^{-1}.$ To reduce uncertainties due to background subtraction, images within 25\dg of the moon and within 2,000 seconds of an annealing event were avoided. To improve the image depth, we also use images from the {\it Neo-WISE} project, which is the extended {\it WISE} mission to find asteroids. We limit the total number of co-added frames to 200 to keep the processing time manageable. The depth of the final cutouts depends on the position of a target relative to the ecliptic plane, with coverage density larger at the ecliptic poles and smaller close to the ecliptic plane \citep{2010AJ....140.1868W}. We used Vega-AB offsets of 2.699 mag for {\it W1} and 3.339 mag for {\it W2} to transform from the Vega to the AB system following Table~3
of Section~IV.4.h of the Explanatory Supplement to the {\it WISE} All-Sky
Data Release Products\footnote{\url{http://wise2.ipac.caltech.edu/docs/release/allsky/expsup/ sec4\_4h.html}}.

\subsection{Photometry} \label{subsec:photometry}

Photometry was performed using the modified version of the photometry pipeline originally developed to generate the {\it WISE} Nearby Galaxy Atlas (WNGA; M. Seibert et al., in preparation). The process starts with taking all geometrical information from the HyperLEDA catalog to define initial photometry apertures which are later modified either by visual inspection or with the aid of {\tt SExtractor} \citep{1996A&AS..117..393B} or {\tt DS9}\footnote{\url{http://ds9.si.edu/site/Home.html}}. We estimate the sky background level within an annulus far from an aperture enclosing the galaxy. All foreground Galactic stars and overlapping objects are masked. The light profile of each galaxy is calculated within concentric elliptical annuli with increments of 3'' for {\it u,g,r,i,z} bands and 6'' for {\it W1} and {\it W2} bands. Resulting light profiles and growth curves are visually inspected and further masking and annulus adjustments are applied if necessary. The entire process is repeated until the growth curves converge.

The sky level of our {\it SDSS} cutouts is minimal because images have been sky-subtracted. On the infrared side, the large resolution elements of {\it W1} and {\it W2} images ($\sim6$'') make it more challenging to estimate accurate sky values. Faint unresolved and therefore unmasked galaxies and foreground stars in the sky annulus influence the quality of photometry results. These issues are addressed by visual inspections as well as control of masking limits. For most galaxies in our sample, we adopt the same recipe explained in \citet{2014ApJ...792..129N} for the photometry of {\it W1} and {\it W2} images and the measurement of the sky level. In addition, for galaxies with a non-convergent curve of growth, iterative micro-adjustments to the background level are evaluated to avoid surface brightness flares or drops and that allow the curve of growth to converge. The adjustment levels are a few percent of the initial estimations of the sky background that were judged within the background annulus.  % No. of frames average: 57 median=31 and stdev=64

For each galaxy, the WNGA pipeline derives radial light profiles and provides two versions of ``total magnitude''. (1) The asymptotic magnitude is calculated within the aperture radius at the point the cumulative luminosity curve of growth flattens. Asymptotic magnitudes are robust since they do not depend on the aperture size chosen by the user. (2) An isophotal magnitude is defined within 25.5 mag arcsec$^{-2}$ then augmented with an extrapolation calculated from the extension of an exponential fit of the galaxy disk to infinity \citep{1996AJ....112.2471T, 2014ApJ...792..129N}. The average discrepancy between these two types of magnitude is at most $0.02$ mag in all bands. The standard deviation of the scatter is $\sim 0.02$ mag for the brightest galaxies, increasing with fainter objects, however on average it remains below $0.05$ mag for all wavebands except the lower quality {\it u}-band ($\sigma \sim 0.1$ mag). Although the results of this study are insensitive to the chosen magnitude, we prefer asymptotic magnitudes to avoid any assumptions about galaxy type. 

For each waveband, $\lambda$, the measured total magnitude of each galaxy, $m^{total}_{\lambda}$, is corrected for obscuration in the Milky Way, redshift with respect to each passband ($k$-correction), and aperture effects, but {\bf not} host obscuration, using the following relation 
\begin{equation}
\label{Eq:bka}
\overline{m}^\lambda = m_{total}^{\lambda} - A^{\lambda}_b - A^{\lambda}_k  - A^{\lambda}_a ~,
\end{equation}
where $A^{\lambda}_b$ is the Milky Way extinction calculated from $R_{\lambda} E(B-V)$, $A^{\lambda}_k$ is the $k$-correction due to Doppler shift and $A^{\lambda}_a$ is the total flux aperture correction. We use \citet{1998ApJ...500..525S} 100 $\mu m$ cirrus maps to extract Milky Way $E(B-V)$ values. For {\it u,g,r,i} and {\it z} bands, $R_{u,g,r,i,z}$ values are given by \citet{2011ApJ...737..103S}. $R_{W1}$ and $R_{W2}$ coefficients are $0.186$ and $0.123$ respectively \citep{1999PASP..111...63F}. The $k$-corrections are very small over the redshift range of interest. 
At optical bands, $k$-corrections are derived following the procedure presented by \citet{2010MNRAS.405.1409C}. For {\it W1} and {\it W2}, $k$-corrections are roughly the same and independent of galaxy type and follow a linear form $A^{W1,2}_k = -2.27z$ \citep{1968ApJ...154...21O, 2007ApJ...664..840H}. Aperture corrections are applied on the {\it WISE} magnitudes where point sources were used for photometric calibration within fixed apertures with the radius of $\sim8.25$'', which is much smaller than our sample galaxies. As a result, the photometry from point source calibrations must be corrected to account for the scattered diffuse light from extended sources. For each galaxy, we apply fixed corrections on {\it W1} and {\it W2} magnitudes, i.e.  $A^{W1}_a=-0.034$ and $A^{W2}_a=-0.041$ mag, taken from Table 5 of Section IV.4.c of the {\it WISE} Explanatory Supplement\footnote{\url{http://wise2.ipac.caltech.edu/docs/release/allsky/expsup/ sec4\_4c.html}}.

% \clearpage
\begin{turnpage}
\tabletypesize{\tiny}
\setlength{\tabcolsep}{0.1cm}
\begin{deluxetable*}{rllllllllllllllllllllll}
\tablecolumns{16}
\tablecaption{Data Catalog\tablenotemark{*}
\label{tab_data}}
%\setlength{\tabcolsep}{0.5em}
%\tablewidth{700pt}
%\setlength{\tabcolsep}{0.01cm}
\tabletypesize{\scriptsize}
\tablehead{ \\
\colhead{PGC} & \colhead{Name} & 
\colhead{$W_{mx}$} & \colhead{$\log (W^i_{mx})$} & \colhead{$m_{21}$} & 
\colhead{$u$} & \colhead{$g$} & \colhead{$r$} & 
\colhead{$i$} & \colhead{$z$} & 
\colhead{$W1$} & \colhead{$W2$} & 
\colhead{$Inc.$} &
\colhead{$R_e$} &
\colhead{$C_{82}$} & \colhead{$b/a$} &
\colhead{$\overline{u}$} & \colhead{$\overline{g}$} & \colhead{$\overline{r}$} & 
\colhead{$\overline{i}$} & \colhead{$\overline{z}$} & 
\colhead{$\overline{W1}$} & \colhead{$\overline{W2}$} \\
\colhead{} & \colhead{} & \colhead{\kms} & \colhead{} &
\colhead{mag} & \colhead{mag} & \colhead{mag} & 
\colhead{mag} & \colhead{mag} & \colhead{mag} & 
\colhead{mag} & \colhead{mag} & \colhead{deg} &
\colhead{arcmin} & 
\colhead{} & \colhead{} & 
\colhead{mag} & \colhead{mag} & \colhead{mag} & 
\colhead{mag} & \colhead{mag} & 
\colhead{mag} & \colhead{mag} \\
\colhead{(1)} & \colhead{(2)} & \colhead{(3)} & 
\colhead{(4)} & \colhead{(5)} & \colhead{(6)} & 
\colhead{(7)} & \colhead{(8)} & \colhead{(9)} & 
\colhead{(10)} & \colhead{(11)} & \colhead{(12)} & 
\colhead{(13)} & \colhead{(14)} &
\colhead{(15)} & \colhead{(16)} &
\colhead{(17)} & \colhead{(18)} & \colhead{(19)} &
\colhead{(20)} & \colhead{(21)} &
\colhead{(22)} & \colhead{(23)}
}
\startdata
4  & PGC000004   &  154$\pm$3  &  2.189$\pm$0.011  &  16.69$\pm$0.08  &  17.57  &  16.43  &  15.91  &  15.57  &  15.42  &  16.06  &  16.42  &  86$\pm$2  &  0.33  &  3.34  &  0.33  &  17.11  &  16.11  &  15.69  &  15.41  &  15.30  &  16.08  &  16.44  \\
16  & PGC000016   &  296$\pm$14  &  2.515$\pm$0.025  &  17.21$\pm$0.18  &  16.08  &  14.83  &  14.21  &  13.86  &  13.61  &  13.98  &  14.67  &  65$\pm$4  &  0.20  &  2.66  &  0.64  &  15.86  &  14.69  &  14.12  &  13.79  &  13.56  &  14.02  &  14.71  \\
55  & UGC12898   &  179$\pm$10  &  2.264$\pm$0.025  &  15.76$\pm$0.08  &  16.82  &  16.08  &  15.71  &  15.50  &  15.40  &  16.18  &  16.60  &  77$\pm$3  &  0.26  &  2.76  &  0.46  &  16.53  &  15.90  &  15.58  &  15.42  &  15.33  &  16.21  &  16.63  \\
68  & ESO538-017   &  206$\pm$18  &  2.395$\pm$0.043  &  16.88$\pm$0.18  &  16.31  &  15.24  &  14.86  &  14.69  &  14.49  &  14.99  &  15.49  &  56$\pm$4  &  0.19  &  3.15  &  0.66  &  16.12  &  15.15  &  14.79  &  14.67  &  14.46  &  15.04  &  15.54  \\
70  & UGC12900   &  432$\pm$2  &  2.636$\pm$0.003  &  15.05$\pm$0.08  &  16.70  &  15.41  &  14.66  &  14.25  &  13.87  &  13.55  &  14.08  &  90$\pm$1  &  0.43  &  3.10  &  0.14  &  16.27  &  15.10  &  14.46  &  14.10  &  13.76  &  13.59  &  14.12  \\
76  &  UGC12901   &  390$\pm$4  &  2.624$\pm$0.013  &  15.78$\pm$0.08  &  15.83  &  14.39  &  13.65  &  13.26  &  13.00  &  13.21  &  13.77  &  68$\pm$4  &  0.32  &  3.57  &  0.47  &  15.52  &  14.18  &  13.51  &  13.16  &  12.93  &  13.25  &  13.82  \\
102  & IC5376   &  427$\pm$6  &  2.635$\pm$0.007  &  15.41$\pm$0.08  &  15.81  &  14.28  &  13.45  &  13.02  &  12.73  &  12.85  &  13.45  &  82$\pm$3  &  0.34  &  4.26  &  0.27  &  15.42  &  14.00  &  13.27  &  12.88  &  12.63  &  12.87  &  13.48  \\
117  & IC1526   &  193$\pm$19  &  2.478$\pm$0.142  &  17.43$\pm$0.08  &  15.86  &  14.60  &  13.95  &  13.58  &  13.37  &  13.45  &  13.97  &        &  0.18  &  2.99  &  0.88  &  15.44  &  14.31  &  13.76  &  13.44  &  13.27  &  13.51  &  14.03  \\
118  & PGC000118   &  113$\pm$6  &  2.248$\pm$0.138  &  16.49$\pm$0.08  &  16.36  &  15.46  &  15.05  &  14.82  &  14.66  &  15.10  &  15.74  &        &  0.23  &  3.32  &  0.94  &  16.07  &  15.28  &  14.92  &  14.75  &  14.60  &  15.14  &  15.78  \\
124  & UGC12913   &  251$\pm$5  &  2.402$\pm$0.010  &  15.78$\pm$0.08  &  17.23  &  15.85  &  15.26  &  14.96  &  14.76  &  15.18  &  15.75  &  85$\pm$2  &  0.34  &  3.41  &  0.29  &  17.04  &  15.74  &  15.18  &  14.91  &  14.72  &  15.22  &  15.79  \\
146  & UGC12916   &  162$\pm$2  &  2.325$\pm$0.027  &  16.72$\pm$0.08  &  16.80  &  15.72  &  15.22  &  14.92  &  14.82  &  15.43  &  16.02  &  50$\pm$4  &  0.27  &  3.14  &  0.74  &  16.59  &  15.60  &  15.13  &  14.87  &  14.78  &  15.47  &  16.06  \\
155  & PGC000155   &  212$\pm$20  &  2.351$\pm$0.042  &  17.60$\pm$0.08  &  16.92  &  15.84  &  15.28  &  14.97  &  14.74  &  15.01  &  15.45  &  71$\pm$3  &  0.17  &  2.48  &  0.60  &  16.44  &  15.51  &  15.05  &  14.82  &  14.62  &  15.05  &  15.50  \\
165  & UGC12920   &  274$\pm$1  &  2.445$\pm$0.005  &  16.22$\pm$0.08  &  16.67  &  15.59  &  15.06  &  14.75  &  14.58  &  14.78  &  15.14  &  81$\pm$3  &  0.24  &  3.06  &  0.47  &  16.37  &  15.39  &  14.92  &  14.66  &  14.51  &  14.83  &  15.19  \\
176  & PGC000176   &  349$\pm$13  &  2.630$\pm$0.027  &  16.08$\pm$0.08  &  15.29  &  14.13  &  13.54  &  13.21  &  13.05  &  13.19  &  13.74  &  55$\pm$4  &  0.24  &  3.19  &  0.60  &  15.03  &  13.96  &  13.43  &  13.13  &  12.99  &  13.23  &  13.78  \\
179  & PGC000179   &  289$\pm$20  &  2.462$\pm$0.030  &  17.09$\pm$0.08  &  16.74  &  15.34  &  14.50  &  14.05  &  13.78  &  13.84  &  14.33  &  86$\pm$2  &  0.25  &  3.44  &  0.27  &  16.38  &  15.08  &  14.33  &  13.92  &  13.69  &  13.87  &  14.36  \\
185  & IC5379   &  133$\pm$20  &  2.167$\pm$0.067  &  17.78$\pm$0.08  &  17.07  &  15.96  &  15.46  &  15.17  &  14.99  &  15.60  &  16.21  &  65$\pm$4  &  0.19  &  2.96  &  0.62  &  16.83  &  15.81  &  15.36  &  15.11  &  14.94  &  15.64  &  16.25  \\
186  & UGC00003   &  472$\pm$4  &  2.707$\pm$0.013  &  15.37$\pm$0.08  &  15.98  &  14.30  &  13.52  &  13.10  &  12.87  &  13.10  &  13.70  &  68$\pm$4  &  0.28  &  3.62  &  0.54  &  15.71  &  14.13  &  13.42  &  13.02  &  12.81  &  13.15  &  13.75  \\
201  & UGC00004   &  302$\pm$19  &  2.572$\pm$0.035  &  17.04$\pm$0.08  &  16.07  &  14.84  &  14.25  &  13.89  &  13.66  &  13.99  &  14.58  &  54$\pm$4  &  0.20  &  3.20  &  0.69  &  15.84  &  14.70  &  14.16  &  13.83  &  13.61  &  14.05  &  14.64  \\
205  & UGC00005   &  446$\pm$20  &  2.697$\pm$0.024  &  15.63$\pm$0.08  &  15.10  &  13.74  &  13.08  &  12.72  &  12.49  &  12.61  &  13.13  &  64$\pm$4  &  0.30  &  2.92  &  0.56  &  14.83  &  13.56  &  12.96  &  12.64  &  12.43  &  12.66  &  13.18  \\
212  & IC5381   &  677$\pm$11  &  2.838$\pm$0.008  &  15.99$\pm$0.08  &  16.16  &  14.55  &  13.70  &  13.23  &  12.92  &  13.04  &  13.68  &  80$\pm$3  &  0.26  &  3.29  &  0.42  &  15.82  &  14.32  &  13.56  &  13.12  &  12.85  &  13.12  &  13.76  \\
\nodata \\
\enddata
\tablenotetext{*}{The complete version of this table is available online.}
\end{deluxetable*}
\end{turnpage}
\clearpage

\subsection{Galaxy Inclinations} \label{subsec:inclination}

The inclinations of spiral galaxies can be coarsely derived from the ellipticity of apertures used for photometry, assuming that the image of a spiral galaxy is the projection of a disk with the shape of an oblate spheroid \citep{2000ApJ...533..744T}. For $\sim 1/3$ of spiral galaxies the approximation of axial ratios provides inclination estimates good to better than 5\dg, with degradation to $\sim 5$\dg for another $1/3$.  However in $\sim 1/3$ of cases, ellipticity-derived inclinations are problematic for a variety of reasons.  Prominent bulges can dominate the axial ratio measurement, with the Sombrero galaxy (NGC 4594) providing an extreme example.  Some galaxies may not be axially symmetric due to tidal effects.  High surface brightness bars within much lower surface brightness disks can lead to large errors.  Simply the orientation of strong spiral features with respect to the tilt axis (spirals opening onto the minor vs the major axis) can be confusing.  The statistical derivation of inclinations for large samples has been unsatisfactory.  With great effort, galaxy images can be broken into multiple components, isolating the disk \citep{2015ApJS..219....4S}.  While such an analysis can give inclinations accurate to $2-3$\dg in clean cases, in many cases, especially in samples of tens of thousands of galaxies with images that degrade with distance, uncertainties can be unsatisfactorily large.  These problems compel a new approach. 

We have investigated the capability of the human eye to evaluate galaxy inclinations. We begin with the advantage that a substantial fraction of spiral inclinations are well defined by axial ratios.  These good cases give us a grid of standards of wide morphological types over the inclination range that particularly interests us of $45-90$\dg.  The challenge is to fit random target galaxies into the grid, thus providing estimates of their inclinations.

To achieve our goal, we designed an online graphical tool, {\it Galaxy Inclination Zoo (GIZ)}\footnote{\url{http://edd.ifa.hawaii.edu/inclination/index.php}}, to measure the inclination of spiral galaxies in our sample. In this graphical interface, we use the colorful images provided by {\it SDSS} as well as the {\it g, r} and {\it i} band images generated for our photometry program. These latter are presented in black-and-white after re-scaling by the $asinh$ function to differentiate more clearly the internal structures of galaxies. The inclination of standard galaxies were initially measured based on their $I$-band axial ratios \citep{2012ApJ...749...78T}.
Each galaxy is compared with the standard galaxies in two steps. First,  the user locates the galaxy among nine standard galaxies sorted by their inclinations ranging between 45\dg and 90\dg in increments of 5 degrees. 
In step two, the same galaxy is compared with nine other standard galaxies whose inclinations are one degree apart and cover the 5\dg interval found in the first step. At the end, the inclination is calculated by averaging the inclinations of the standard galaxies on the left/right-side of the target galaxy. In the first step, if a galaxy is classified to be more face-on than 45\dg, it is flagged and step two is skipped.

Initially, the tool was tested by two of the authors (EK looked at essentially all of the candidates and RBT gave attention to about half).  The tool was further tested with the participation of 10 undergraduate students and friends, with multiple cross-checks allowing the assessment of consistency and potentially revealing systematic differences.
The program was then opened to the public and hence benefited from the participation of citizen scientists and tens of amateur astronomers. Ultimately, for each spiral we use all the measured inclinations by all users who worked on that galaxy.

We take the following precautions to minimize user dependant and independent biases. (1) We round the resulting inclinations to the next highest or smallest integer values chosen randomly. (2) At each step, standard galaxies are randomly drawn with an option for users to change them randomly to verify their work or to compare galaxies with similar structures. (3) To increase the accuracy of the results, we catalog the median of at least three different measurements performed by different users. (4) Users may reject galaxies for various reasons and leave comments with the aim of avoiding dubious cases.

The uncertainties on the measured inclinations are estimated based on the statistical scatter in the reported values by different users.  We believe that we have achieved a statistical accuracy of $\pm 4$\dg rms. A more detailed discussions of these measurements and their uncertainties will be given in an upcoming data paper (Kourkchi et al. 2019, in preparation).

An obvious step forward would be to replace the human eye with a machine learning algorithm.  Such an endeavor would require an instruction set of order $10^4$ representative galaxies with reasonably established inclinations.  Our entire sample is of comparable size to the required training set.  Machine learning will be called for with larger samples in the future.

\subsection{Data Catalog} \label{subsec:catalog}

The results of our measurements and the collection of other parameters for 2,239 spiral galaxies we used in this study are gathered in Table \ref{tab_data}. Columns of this table include (1) Principal Galaxies Catalog (PGC) number and (2) common name. (3) is \hi linewidth in \kms and (4) is logarithm of the inclination-corrected \hi linewidth, calculated from $W^i_{mx}=W_{mx}/\sin(i)$, where $i$ is the inclination angle presented in column (13). Column (5) is \hi 21cm magnitude calculated from the \hi flux, $F_{HI}$, using $m_{21}=-2.5 \log F_{HI}+17.4$. Columns (6-10) are the {\it SDSS u,g,r,i} and {\it z} total magnitudes in the 
AB system. Columns (11-12) are {\it WISE} {\it W1} and {\it W2} total magnitudes in the AB system. Except for the {\it u}-band, the uncertainties on all the measured magnitudes are not worse than $0.05$ mag, the value we adopt conservatively for error propagation in this study. For the {\it u}-band magnitudes, uncertainties are $\sim 0.1$ mag. Column (13) stores our measured inclinations, with empty cells for galaxies with inclinations less than 45\dg (see \S\ref{subsec:inclination}). Columns (14) is the half light radii measured at $W2$ bands. Column (15) holds the concentration index that is defined as $C_{82}=5{\rm log}(r_{80}/r_{20})$, where $r_{80}$ and $r_{20}$ are the semi-major axes of the isophots enclosing $80\%$ and $20\%$ respectively of the total galaxy light at $W2$ band \citep{1985ApJS...59..115K}. Column (16) contains the axial ratio, $b/a$, of the elliptical apertures that we used for the photometry of $W1(2)$ band images, where $a$ and $b$ are the aperture semi-major and semi-minor axes. Columns (17-23) tabulate optical/infrared magnitudes corrected for Milky Way obscuration, redshift $k-$correction, and aperture effects, but {\bf not} host inclination, following Eq. \ref{Eq:bka}, based on the corresponding raw magnitudes listed in columns (6-12).

\section{Intrinsic Obscuration in Spiral Galaxies} \label{sec:iog}

The observed light from the stellar content of a galaxy depends on interactions with its own Inter-Stellar Medium (ISM). The level of flux attenuation depends on the light wavelength and the physical nature of the ISM, as well as the orientation of galaxy relative to the observer line-of-sight. The composition of the ISM depends on the type, mass, age, metallicity, and the evolutionary history of a galaxy \citep{2011EP&S...63.1027I}. Galaxy light is altered through the absorption of ultraviolet/visible photons by the ISM dust particles depending on dust grain size and emission of the received energy at longer wavelengths. Increasing the galaxy inclination, photons have a longer path-length through the ISM, and therefore the obscuration effect is more pronounced. Accordingly, it is crucial to correct for obscuration when we use the TFR to measure distances of galaxies that are preferably chosen to have larger inclinations. 
There is the trade-off that kinematic errors diminish toward edge-on orientation while photometric errors increase.  As a practical matter, it is more important to control the kinematic errors.

Understanding all physical processes involved in the intrinsic extinction of galaxies requires a much more detailed analysis that is beyond the scope of this study. Here, our main goal is to empirically estimate the inclination dependent part of dust obscuration in spirals by correlating the obscuration with various distance independent observable parameters representing different properties of galaxies. We list the observable parameters in what follows. (1) The \hi linewidth corrected for inclination, $\log (W^i_{mx})$, is related to the absolute magnitude of a galaxy via the TFR and therefore is a good proxy for the galaxy mass/size. (2) The optical-infrared colors ($\overline{m}_\lambda - \overline{W}j$), characterizes the attenuation of optical fluxes through the known property that the dust obscuration diminishes as wavelengths increases until it is ultimately very tiny or negligible at infrared bands. (3) Another useful parameter is the pseudo-color difference between 21 cm and infrared magnitudes that probes \hi to stellar flux, defined as \begin{equation}
C_{21Wj}=m_{21}-\overline{W}j,
\end{equation}
where $m_{21}$ is the \hi 21~cm magnitude and $\overline{W}j$ is either the $\overline{W}1$ or the $\overline{W}2$ magnitude. Since $C_{21Wj}$ is constructed based on fluxes at infrared and longer wavelengths, it should be free of obscuration effects and an accurate proxy for the type of galaxy. In the absence of many active star forming regions, galaxies with more negative (or bluer) $C_{21Wj}$ colors have larger \hi to stellar mass ratio and vice versa. There is the caveat that in large star forming spirals the infrared emission from hot dust near the star forming regions and stochastic heating of small grains in diffuse regions augment the observed infrared fluxes \citep{2018ApJS..236...32L}. In spirals with many star forming regions, $W_j$ might be a poor proxy for stellar mass content such that $C_{21Wj}$ underestimates the ratio of \hi to stellar mass. (4) The average surface brightness of galaxies is another distance independent parameter that characterizes the galaxy type and its morphology. In this study, we use the effective surface brightness parameter calculated within the half light radius in $\overline{W}1/\overline{W}2$ infrared bands, $\langle \mu_j \rangle_e$. We correct the effective surface brightness to compensate for the effect of galaxy inclination using \begin{equation}
\langle \mu_j \rangle^{(i)}_e=\langle \mu_j \rangle_e+0.5 \log (a/b), 
\end{equation}
where $a/b$ is the axial-ratio of the elliptical aperture used for photometry. (5) The galaxy light concentration index, $C_{82}$, is another distance independent photometry product that encodes bulge vs. disk morphological information that might be useful in our analysis.  

\subsection{Fiducial correlations: face-on spirals} \label{sec:fcs}

\begin{figure}[ht]
\centering
\includegraphics[width=0.89\linewidth]{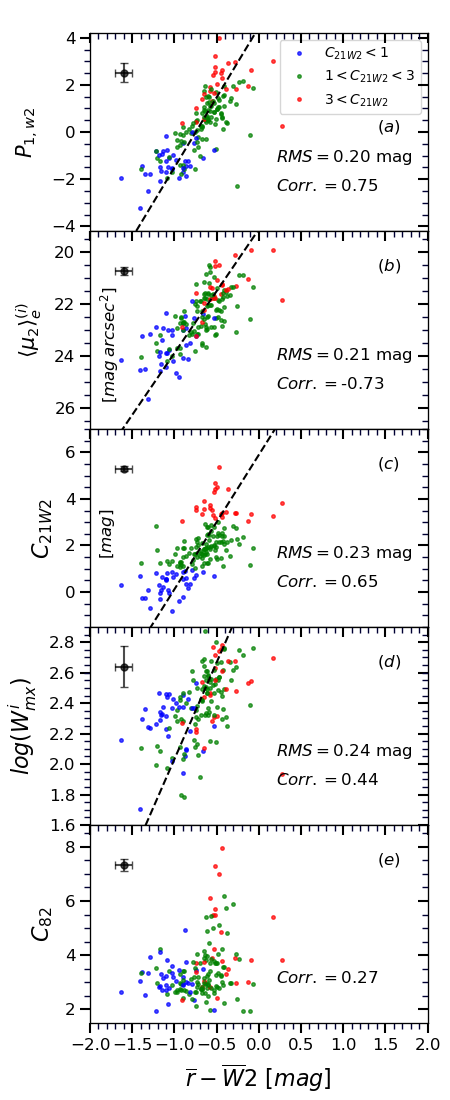}
\caption{$\overline{r} - \overline{W}2$ vs. distance independent observables for galaxies more face-on than 45$^{\circ}.$ Dashed lines represent the best linear fit with minimizing residuals along the horizontal axis. Each point represents a galaxy with blue, green and red colors corresponding to $C_{21W2}<1$, $1<C_{21W2}<3$ and $C_{21W2}>3$ respectively. In each panel, $Corr.$ is the correlation factor for plotted parameters.}
\label{fig:r_w2_features_Fon}
\end{figure}

\begin{figure}[ht]
\centering
\includegraphics[width=1.\linewidth]{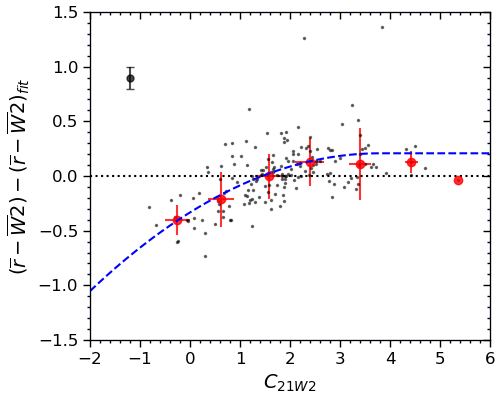}
\caption{Deviation of $\overline{r}-\overline{W}2$ colors from the best fitted line in panel (d) of Fig. \ref{fig:r_w2_features_Fon} versus the HI$-$IR flux pseudo-color parameter, $C_{21W2}$. Each black dot represent an individual galaxy. Red points show the median position of black points within $C_{21W2}$ bins each with the size of 1 mag. The error bars on the red points show the 1-$\sigma$ scatter of the black points within each bin. The black error bar on the top left corner shows the typical error bar of the black dots. The blue dashed curve displays the best fitted function to the black points. This function has a quadratic form on the left side and continues as a flat horizontal line after it reaches its maximum. The fitted quadratic function deviates from the null value as $-0.038~C_{21W2}^2+0.285~C_{21W2}-0.333$.}
\label{fig:deviation_c21w2}
\end{figure}

Prior to examining the inclination dependent part of attenuation, we first need to establish inclination-free fiducial correlations between optical-infrared colors and other distance independent observables. Then we can attribute any deviations from the fiducial relations to the inclination of the galaxies. To establish the fiducial relations, in this section we restrict our sample to the more face-on spirals where the effect of inclination on attenuation will be at a minimum. To this end, we give attention to the 225 spirals in our sample that were flagged to be more face-on than 45\dg by our more accurate inclination determination (\S \ref{subsec:inclination}). We arbitrarily set the inclination to be $40\pm5$ degrees for all of these galaxies to be able to perform our analysis in this section and to study the useful strong correlations between different observables. Although the inclinations of these nearly face-on galaxies are not exactly zero, their obscuration relative to face-on is minimal. Once we formulate the empirical fiducial model for face-on spirals, we no longer use these 225 nearly face-on galaxies in the numerical optimization of our model.

Fig.~ \ref{fig:r_w2_features_Fon} illustrates the correlation between $\overline{r}-\overline{W}2$ color and different observable features introduced above for the 225 nearly face-on galaxies. Each point in this diagram represents a galaxy, color coded based on the parameter $C_{21W2}$; blue to red proceeding from gas prevalent to gas depleted. The typical error bar of the scattered points is given at the top left corner of each panel. The RMS is the root mean squared of deviations from the fitted dashed lines along the horizontal axis. Panels of this diagram are sorted based on the correlations between the plotted parameters, with the bottom panel (e) showing the weakest correlation between color and concentration parameters.

Panel (d) displays the correlation between color and $\log (W^i_{mx})$. In this plot, the relative position of points and hence our final conclusions are insensitive to the arbitrarily chosen 40$\pm$5 degrees for the inclination when estimating $\log (W^i_{mx})$. True values of inclinations statistically shuffle points vertically on the $\log (W^i_{mx})$ axis and only weakly affects the general correlation. The correlation between $\log (W^i_{mx})$ and color is related to a color-magnitude correlation given that more massive, brighter galaxies have larger $\log (W^i_{mx})$ values. Larger galaxies tend to have older mean stellar populations and as a result their optical-infrared colors are redder.

The observed correlation in panel (d) is not very strong (correlation coefficient equals $0.44$). We are motivated to explore the possible role of other parameters to explain the scatter of galaxies around the fitted straight line. It is seen that almost all red points are distributed on the right side of the fitted line and most blue points are on its left side in panel (d). The red points represent galaxies with more positive $C_{21W2}$ that have less \hi gas relative to their stellar content.  These galaxies have more completely converted their gas into stars endowing them with older stellar populations and redder optical-infrared colors. 

To further investigate this additional correlation, in Fig. \ref{fig:deviation_c21w2} we plot the horizontal deviation of galaxies from the fitted line in panel (d) of Fig.~ \ref{fig:r_w2_features_Fon} versus the $C_{21W2}$ colors. In this figure, each black point represents a galaxy and red points show the average position of galaxies in successive $C_{21W2}$ bins. Galaxies with more positive $C_{21W2}$ values deviate towards redder $\overline{r}-\overline{W}2$ values. There are few spirals with $C_{21W2}>4$ {\rm mag}, but it appears that the trend of deviations from the fiducial line levels off around $\overline{r}-\overline{W}2 \sim 3$.

Panel (c) of Fig. \ref{fig:r_w2_features_Fon} shows the reasonably strong correlation between $\overline{r}-\overline{W}2$ and $C_{21W2}$ colors.  The larger correlation coefficient of 0.65, compared to that of panel (d), 0.44, suggests that $C_{21W2}$ carries greater information about the optical-infrared parameter. There is a similar correlation with deviations from the fitted line in plot (c) to that with the $\log (W^i_{mx})$ parameter in plot (d).

The correlation between the effective surface brightness, $\langle \mu_2 \rangle^{(i)}_e$ measured in $\overline{W}2$-band and $\overline{r}-\overline{W}2$ is displayed in the panel (b) of Fig. \ref{fig:r_w2_features_Fon}. Surprisingly this correlation is even tighter than those presented in the panels (c) and (d). 

In conclusion, larger more intrinsically luminous galaxies have redder stellar population in general, they tend to have higher surface brightness, and lesser \hi to stellar mass fractions. 

\subsection{Distance independent Principal Components} \label{sec:pca}

\begin{figure*}[ht]
\centering
\includegraphics[width=1.\linewidth]{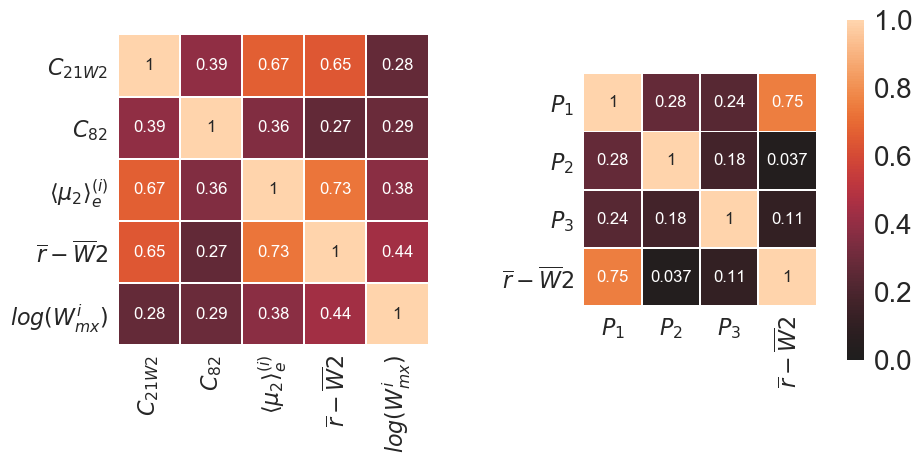}
\caption{Correlation matrices for distance independent observables for spirals more face-on than 45\dg where the absolute value of correlation coefficients are reported. }
\label{fig:corr_table}
\end{figure*}

In \S \ref{sec:fcs}, we presented Fig.~\ref{fig:deviation_c21w2} as an example to show how using the additional correlations that exist between different parameters help to better understand the fiducial colors of face-on spirals. In fact, all the discussed observable features in \S \ref{sec:fcs} are correlated with one another. The left panel of Fig. \ref{fig:corr_table} shows the correlation matrix of all features for spirals more face-on than 45\dg. This matrix holds all the correlation coefficients. The correlation of the color parameter, $\overline{r}-\overline{W}2$, with other features was presented in detail in Fig. \ref{fig:r_w2_features_Fon} and \ref{fig:deviation_c21w2}. One of the strongest correlations exists between $C_{21W2}$ and $\langle \mu_2 \rangle^{(i)}_e$ with the factor of $-0.67$. We find $\log (W^i_{mx})$ is mildly correlated to the other features. According to Fig. \ref{fig:corr_table}, $C_{82}$ and $\log (W^i_{mx})$ have similar behaviors in terms of correlation factors, however $C_{82}$ has the smallest correlation with the optical-infrared color term. 
Repeating the same analysis with and without the concentration parameter does not significantly change the outcomes. As such, we decided to remove the concentration parameter from our analysis due to its small correlations and to avoid distracting our model by incorporating a substantially sub-dominant feature. 

\begin{table}[ht]
\setlength{\tabcolsep}{0.1cm}
\centering
\caption{Principal Components and standardization factors defined in Eq. \ref{eq:pca_combo} and \ref{eq:pca_features} using $W2$-band photometric data.} 
\label{tab:pca_w2}
\begin{tabular}{c|ccc|c||cc}
\tablewidth{0pt}
\hline \hline
 \multirow{2}{*}{W2} & \multirow{2}{*}{$p_1$} & \multirow{2}{*}{$p_2$} & \multirow{2}{*}{$p_3$} & Variance & \multirow{2}{*}{$u$} & \multirow{2}{*}{$\sigma$}\\
    &  &   & & Ratio   \\
\hline 
\decimals
$P_1$ & 0.524 & 0.601 & -0.603 & 0.70 & 2.47 & 0.18 \\
$P_2$ & -0.851 & 0.384 & -0.358 & 0.19 & 1.63 & 1.15 \\
$P_3$ & 0.016 & 0.701 & 0.713 & 0.11 & 23.35 & 1.38 \\
\hline
\end{tabular}
\end{table}

Due to the mutual correlation of the discussed observable features, and in order to only use the genuine information of each parameter, we perform a Principal Component Analysis (PCA) using $\log (W^i_{mx})$, $C_{21W2}$ and $\langle \mu_j \rangle^{(i)}_e$ parameters. 
PCA outputs an orthogonal transformation that linearly converts the correlated features to uncorrelated principal components. Using smaller number of principal components compared to the original number of features decreases the sensitivity of results to the uncertainty of measured features by taking advantage of the correlation between features. In fact, one can reduce the noise level of the measured parameters through reducing the number of dimensions by mapping from the input physical features to a space generated by the first few significant principal components followed by an inverse transformation back to the physical space.

Principal components can be derived by linearly combining the observable features following this form
\begin{equation}
\label{eq:pca_combo}
P_{m} = \sum_{i=1}^{3} p_{m,i} ~X_{i},
\end{equation}
where $p_{m,i}$ are weight numbers and $X_i$ is the standardized form of a feature $x_i$ derived using $X_i=(x_i-u_i)/\sigma_i$ with $u_i$ and $\sigma_i$ being the mean and standard deviation of the corresponding feature for all sample galaxies with inclinations larger than 45\dg. These features are

\begin{equation}
\label{eq:pca_features}
  \begin{cases}
    x_{1} = \log (W^i_{mx}) ~,\\
    x_{2} = C_{21Wj} ~,\\
    x_{3} = \langle \mu_j \rangle^{(i)}_e ~.
  \end{cases}
\end{equation}

In Eq. \ref{eq:pca_features}, $\log (W^i_{mx})$ and $\langle \mu_j \rangle^{(i)}_e$ are already corrected for the effect of inclination and $C_{21Wj}$ is free of the inclination dependent dust obscuration. Accordingly, for the sake of accuracy, in this part of our analysis we draw our attention to 2014 galaxies in our sample with inclinations larger than 45\dg where we have measured their inclinations and used this information to calculate the inclination corrected parameters. The principal component coefficients defined in Eq. \ref{eq:pca_combo} and \ref{eq:pca_features} together with the normalization factors, $u_i$ and $\sigma_i$, are presented in Tables \ref{tab:pca_w2} and \ref{tab:pca_w1}. 
One of the properties of principal components is that the main principal component, $P_1$, has the largest variance compared to other components and the input features and therefore it holds the most information on the scatter of data points. The variance ratio of each principal component is defined as the ratio of its variance by the sum of variances of all individual principal components. The variance ratios are often used to evaluate the significance of principal components. As presented in both Tables \ref{tab:pca_w2} and \ref{tab:pca_w1}, $P_1$ carries about $70\%$ of the information on the scatter in the feature space defined in Eq. \ref{eq:pca_features}.

\begin{table}[ht]
\setlength{\tabcolsep}{0.1cm}
\centering
\caption{The same as Table \ref{tab:pca_w2} but for $W1$ band.} 
\label{tab:pca_w1}
\begin{tabular}{c|ccc|c||cc}
\tablewidth{0pt}
\hline \hline
 \multirow{2}{*}{W1} & \multirow{2}{*}{$p_1$} & \multirow{2}{*}{$p_2$} & \multirow{2}{*}{$p_3$} & Variance & \multirow{2}{*}{$u$} & \multirow{2}{*}{$\sigma$}\\
    &  &   & & Ratio   \\
\hline 
\decimals
$P_1$ & 0.533 & 0.595 & -0.602 & 0.71 & 2.47 & 0.18 \\
$P_2$ & -0.844 & 0.420 & -0.332 & 0.18 & 2.11 & 1.17 \\
$P_3$ & 0.055 & 0.685 & 0.726 & 0.11 & 22.83 & 1.37 \\
\hline
\end{tabular}
\end{table}

We evaluate the performance of the derived principal components for the nearly face-on spirals, where the effect of inclination dependent dust attenuation on their optical-infrared colors is less significant. In the right panel of Fig. \ref{fig:corr_table}, the correlation matrix of principal components and $\overline{r}-\overline{W}2$ color of nearly face-on galaxies is presented. As expected, principal components are minimally correlated with one another. Considering both panels of Fig. \ref{fig:corr_table}, the largest correlation exists between $\overline{r}-\overline{W}2$ and the main components, $P_1$, with the correlation factor of $0.75$. Consistently, panel (a) of Fig. \ref{fig:r_w2_features_Fon} also shows how using $P_1$ reduces the scatter and establishes a tighter correlation that can serve as a basis for a fiducial relationship.

To summarize, we found that we can replace all correlated parameters in our problem with the first principal component, $P_1$, a linear combination in roughly equal parts of linewidth, ratio of HI to old stars, and surface brightness
\begin{equation}
\label{Eq:P1}
\begin{split}
    P_{1,W2} = 0.524({\rm log}W^i_{mx}-2.47)/0.18 \\
    + 0.601(C_{21W2}-1.63)/1.15 \\
    - 0.603(\langle \mu_2 \rangle^{(i)}_e-23.35)/1.38
    \end{split}
\end{equation}
that encodes $\sim70\%$ of the information on the scatter of data points in the feature space. It is also easier to use only one principal component compared to several features to describe the scatter and general linear trends we observe in Fig. \ref{fig:r_w2_features_Fon} and \ref{fig:deviation_c21w2}. In addition, any further analysis based on $P_1$ is more robust and less prone to outliers and uncertainties in the measured parameters.
 
The principal components that are alternatively derived based on $W1$ and $W2$ fluxes are subject to different levels of dust obscuration and therefore $P_{1,W1}\neq P_{1,W2}$. In appendix \ref{sec:w1/w2}, we compare these principal components and present a procedure to avoid the inclination dependent biases introduced by using $P_{1,W1}$.

\section{Method} \label{sec:method}

In \S \ref{sec:iog}, we studied the fiducial relation between the optical-infrared color of face-on spirals and their other properties where their luminosities are not influenced by their orientation relative to observer. This study led us to choose the most significant inclination independent/corrected parameters for the analysis of our entire sample of spirals. Finally, the principal component analysis of these observable features in infrared and 21cm wavelength provided us with a novel parameter, $P_1$, which can serve as a proxy for the type of spirals by linearly combining their luminosity (or size), surface brightness, and the fraction of \hi gas to stellar content.

In this section and an appendix, we take two different approaches to empirically model the inclination dependent attenuation in spiral galaxies. In \S \ref{sec:parametrization}, considering the studied fiducial relationship, we build a parametric model for the dust attenuation in spirals. In \S \ref{sec:opt} we sample the posterior distribution of our model parameters using the Markov Chain Monte Carlo method. In addition, we tackle the same problem with a non-parametric model based on a Gaussian Process (GP) formalism in appendix \S \ref{sec:gpc}.

\subsection{Parametric model} \label{sec:parametrization}

Intuitively, the optical-infrared color of a spiral can be modelled as follows by combining its face-on fiducial color derived from its properties and an extra term responsible for any alteration due to its spatial orientation 
\begin{equation}
\label{Eq:mWj}
\overline{m}_\lambda - \overline{W}j = \Delta m_{\lambda J}^{(o)}+A^{(i)}_{\lambda J} ~,
\end{equation}
where $\Delta m_{\lambda J}^{(o)}$ is the fiducial color of a galaxy seen face-on.  The effect of inclination angle would be then encoded in $A^{(i)}_{\lambda J}$ which depends on the galaxy properties and its deviation from face-on. The linear fiducial relation we found in \S \ref{sec:fcs} that relates the main principal component of a spiral galaxy and its face-on color suggests the following linear relation
\begin{equation}
\label{Eq:GlWjO}
\Delta m^{(o)}_{\lambda J} = \alpha_{\lambda J} P_{1,J}+\beta_{\lambda J}
\end{equation}
where $P_{1J}$ is the first principal component derived using Eq. \ref{eq:pca_combo} based on the $\overline{W}j$ band photometry. To be consistent with the standard empirical formalism adopted in the earlier study by \citet{1998AJ....115.2264T}, we assume that $A^{(i)}_{\lambda J}$ is separable and has the form
\begin{equation}
\label{Eq:AWj}
A^{(i)}_{\lambda J} = \gamma_{\lambda J} \mathcal{F}_\lambda(i) ~,
\end{equation}
where $\gamma_{\lambda J}$ is a function of the main principal component and $\mathcal{F}$ depends on the galaxy inclination. We adopt $\mathcal{F}$ to be related to inclination $i$, through
\begin{equation}
\label{Eq:Fli}
\mathcal{F}_\lambda(i)={\rm log}\Big[\cos ^2(i)+q_\lambda^2\sin^2(i)\Big]^{-1/2} ~,
\end{equation}
where $i$ is the inclination angle from face-on. $\mathcal{F_\lambda}$ is zero for face-on galaxies and increases with inclination to capture the dust attenuation effect along the line-of-sight path length. This relation is very similar to the relation that uses the axial ratio of the photometry aperture, $b/a$, with the form of $\mathcal{F}=\log (a/b)$ \citep{1991rc3..book.....D, 1995A&A...296...64B,1998AJ....115.2264T}, where axial ratio is connected to inclination through ${\rm cos}^2~i=[(b/a)^2-q_o^2]/(1-q_o^2)$ with $q_o$ being the edge-on ellipticity of the spiral disk modelled as a prolate ellipsoid. In our model, $q_\lambda$ has a similarity to $q_o$, however it might be interpreted differently. One could think of $q_\lambda$ as a wavelength dependent hyper-parameter that however does not have any direct geometric meaning. For an edge-on spiral ($i=90$\dg) $q_{\lambda}$ sets the peak of $\mathcal{F_\lambda}$ and defines the sensitivity of $\mathcal{F_\lambda}$ to the galaxy orientation in different wavebands. 

Exploring different functional forms, we adopt a third degree polynomial to model the dependency of $\gamma_{\lambda J}$ on the main principal component of spirals, $P_{1,J}$, that encapsulates the necessary information on galaxy properties. We parameterized this relation as 
\begin{equation}
\label{Eq:GlWjI}
\gamma_{\lambda J} = \sum_{n=0}^{3} C^{(n)}_{\lambda J} ~ P^{^n}_{1,J} ~,
\end{equation}
where $C^{(n)}_{\lambda J}$ are the constant coefficients of the adopted polynomial function and $n$ is the degree of each individual term.

\begin{figure*}[t]
\includegraphics[width=\linewidth]{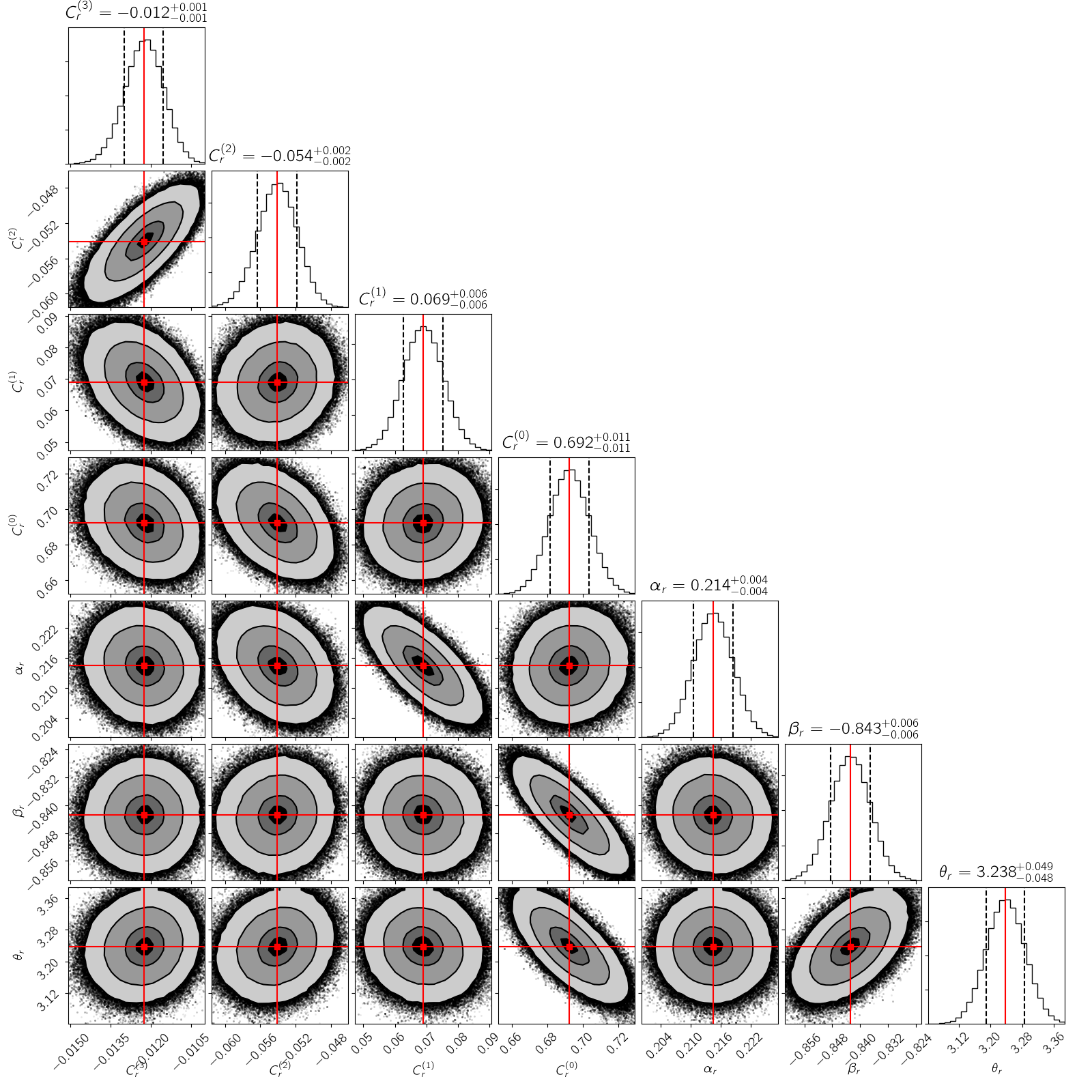}
\caption{The posterior distribution of the optimized parameters to estimate $\overline{r} - \overline{W}2 = \gamma^{(o)}_{r,W2}+A^{(i)}_{r,W2}$. Contours represent $\sigma/2$, $\sigma$, $3\sigma/2$ and $2\sigma$ levels of the 2-dimensional distributions and they enclose 12\%, 39\%, 68\% and 86\% of the distributed points respectively. To facilitate the calculations we set $q_r^2 =10^{-\theta_r}$.}
\label{fig:PC0_mcmc_r_w2}
\end{figure*}

\begin{figure*}[t]
\includegraphics[width=\linewidth]{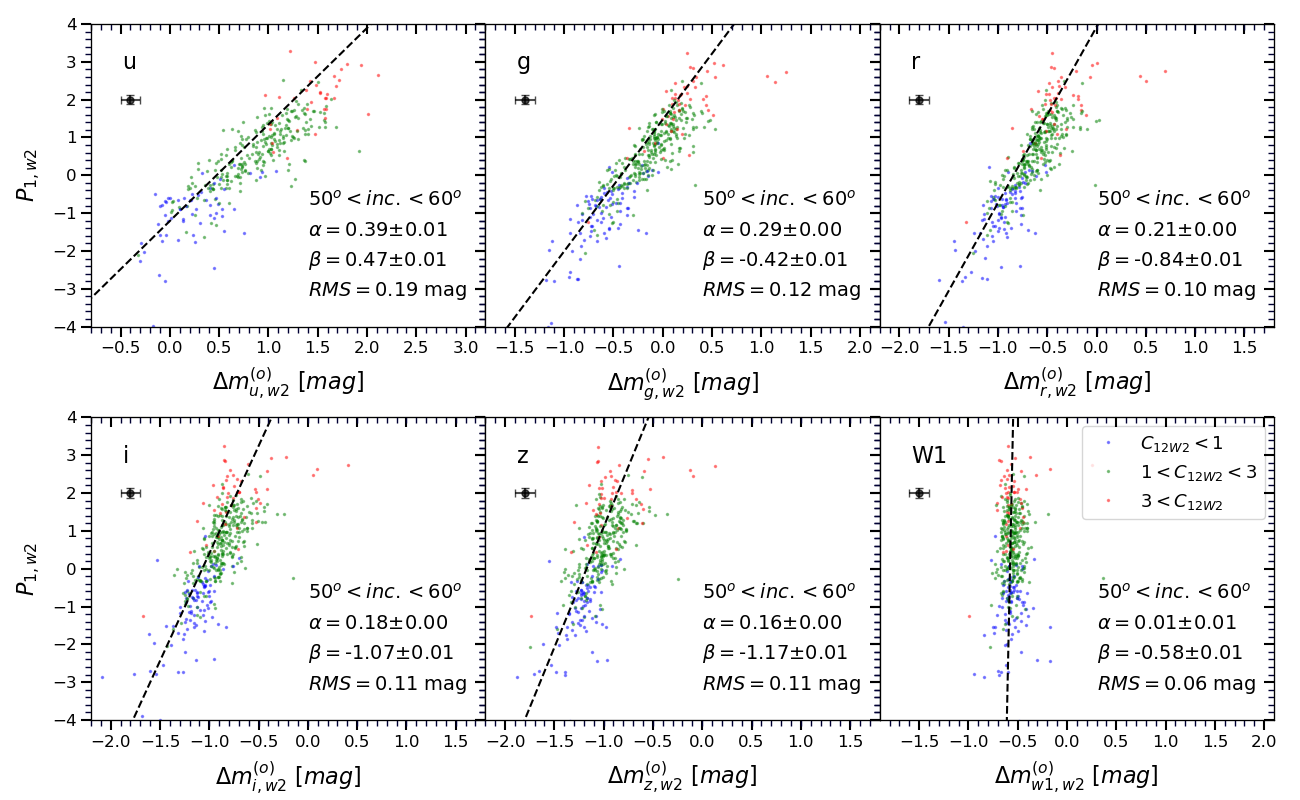}
\caption{First principal component constructed based on the observable features (i.e. $\log (W^i_{mx})$, $C_{21Wj}$ and $\langle \mu_j \rangle^{(i)}_e$) versus the inclination corrected color terms, $\Delta m^{(o)}_{\lambda J}=(\overline{m}_\lambda - \overline{W}2)-A^{(i)}_{\lambda,W2}=\alpha_\lambda P_{1,W2}+\beta_\lambda$. Each point represents a galaxy whose inclination is between 50\dg and 60\dg and color codes are the same as in Fig. \ref{fig:r_w2_features_Fon}. RMS is the root mean squared of the horizontal deviations from the dashed line characterizing face-on systems that is drawn based on the optimized parameters listed in Table \ref{table:model_params}.}
\label{fig:color_pc0_w2_50-60}
\end{figure*}

\begin{deluxetable*}{lccccccc}
\tablecolumns{20}
\tablecaption{Optimized parameters of the parametric model to describe $\overline{m}_\lambda - \overline{W}2$ defined by Eq. \ref{Eq:mWj}.
\label{table:model_params}}
\tabletypesize{\scriptsize} 
\tablehead{\\
\colhead{$\lambda$} &
\colhead{$C^{(3)}_{\lambda}$} & \colhead{$C^{(2)}_{\lambda}$} &  
\colhead{$C^{(1)}_{\lambda}$} & \colhead{$C^{(0)}_{\lambda}$} & 
\colhead{$\alpha_\lambda$} & \colhead{$\beta_\lambda$} &
\colhead{$\theta_\lambda=-2\log (q_\lambda)$} 
}
\startdata
$u$ & $-0.013\pm0.001$ & $-0.055\pm0.003$ & $0.096\pm0.009$ & $1.202\pm0.018$ & $0.392\pm0.006$ & $0.473\pm0.008$ &$2.935\pm0.041$ \\
$g$ & $-0.013\pm0.001$ & $-0.056\pm0.002$ & $0.081\pm0.007$ & $0.873\pm0.013$ & $0.287\pm0.004$ & $-0.424\pm0.006$ &$3.131\pm0.042$ \\
$r$ & $-0.012\pm0.001$ & $-0.054\pm0.002$ & $0.069\pm0.006$ & $0.693\pm0.011$ & $0.214\pm0.004$ & $-0.843\pm0.006$ &$3.237\pm0.049$ \\
$i$ & $-0.012\pm0.001$ & $-0.055\pm0.003$ & $0.051\pm0.007$ & $0.586\pm0.010$ & $0.176\pm0.004$ & $-1.071\pm0.005$ &$3.286\pm0.054$ \\
$z$ & $-0.011\pm0.001$ & $-0.055\pm0.003$ & $0.023\pm0.007$ & $0.474\pm0.010$ & $0.157\pm0.004$ & $-1.173\pm0.005$ &$3.411\pm0.063$ \\
$W1$ & $-0.002\pm0.000$ & $-0.007\pm0.000$ & $0.015\pm0.011$ & $0.062\pm0.018$ & $0.008\pm0.005$ & $-0.578\pm0.006$ & $1.789\pm0.068$\\
\enddata
\end{deluxetable*}

\subsection{Optimization} \label{sec:opt}

To find the best parameters of the model introduced in \S \ref{sec:parametrization}, we follow a Bayesian approach. The objective is to find the posterior probability distribution $\Pc(\Theta|\mathcal{D})$, where $\Theta$ is the vector of all model parameters (i.e. $\alpha_{\lambda J}$, $\beta_{\lambda J}$, $C^{(n)}_{\lambda J}$ and $q_\lambda$). $\Dc$ holds the observed data, i.e. $P_1$ and inclination, for all spirals with inclinations greater than 45\dg. Conditional probability law indicates that $\Pc(\Theta|\mathcal{D}) \propto \Pc(\mathcal{D}|\Theta)\Pc(\Theta)$.

The lack of any prior knowledge on the distribution of the model parameters leads us to set $\Pc(\Theta)=1$. This assumption simplifies our problem which could be solved adopting any optimization technique to maximizes the likelihood function, defined as $\Lc=\Pc(\Dc|\Theta)$. Nevertheless, we take the advantage of the Markov Chain Monte Carlo (MCMC) method to explore the parameter space and to sample the posterior distribution by incorporating all uncertainties of the observed data. The independence of each data point associated to a galaxy from the measured parameters of other galaxies, and the Gaussian nature of uncertainties results in the following likelihood function
\begin{equation}
\label{Eq:likelihood}
\Lc= \prod_{n=1}^{N} \frac{1}{\sqrt{2 \pi \sigma_n^2}} ~\exp\Big(\frac{\Dc_n-\Mc_n(\Theta)}{\sigma_n}  \Big)^2 ~,
\end{equation}
where $n$ is the galaxy index and $N$ is the total number of sample galaxies. For each galaxy $\Dc_n=(\overline{m}_\lambda - \overline{W}j)_n$ is calculated using the observed values of the optical and infrared magnitudes and $\Mc_n$ is the output of the parametric model described by Eq. \ref{Eq:mWj}, and $\sigma_n$ is the uncertainty of the $\Dc_n-\mathcal{M}_n(\Theta)$ parameter that acts as a weight factor that penalizes a model based on its prediction from real measurements and is calculated using $\sigma_n^2=\sigma_{D_n}^2+\sigma_{\Mc_n}^2$. One can obtain $\sigma_{D_n}$ by adding the uncertainties of the measured magnitudes in quadrature. For any model with given parameters, $\Theta$, the model uncertainty $\sigma_{\Mc}$ is calculated through propagating uncertainties on observable features, i.e. $\sigma_{P_{1}}$ and $\sigma_i$.

We use the Python package {\it emcee} \citep{2013PASP..125..306F} to sample the posterior distribution. To accelerate calculation by utilizing fast linear algebra routines, {\it emcee} uses the logarithm of the posterior likelihood 
\begin{equation}
\label{Eq:likelihood1}
\log \Lc({\bf r})= -\frac{1}{2}{\bf r}^T\Sigma^{-1}{\bf r}-\frac{1}{2}\log |\Sigma|-\frac{N}{2}\log (2 \pi) ,
\end{equation}
where ${\bf r}$ is the $N\times1$ data-model residual vector whose elements are $r_n = \Dc_n-\mathcal{M}_n(\Theta)$ and $\Sigma$ is the $N\times N$ diagonal covariance matrix where $\Sigma_{n,n}=\sigma^2_n$. Adopting this likelihood function, for each photometry band we generate 200 chains each with the length of 20,000 samples. To expand the size of explored regions in the parameter space, each chain is initialized randomly. We observed that after almost 500 burning steps, each chain converges and follows the Markov statistics. To be more conservative, we removed the first 2,000 steps of each MCMC sample and combined all 200 walkers to construct the posterior distribution of model parameters. Fig. \ref{fig:PC0_mcmc_r_w2} illustrates the corner plot for the distribution of the sampled distribution of parameters to model $\overline{r}-\overline{W}2$. In this diagram, the top-most panel of each column shows the one dimensional distribution of the corresponding sampled parameter, displayed with the median values (red solid line) and the lower/upper bounds corresponding to 16/84 percentiles (black dashed line). Horizontal and vertical red lines show the location of the median values which we adopt as the optimum values of our model parameters. The shape of the posterior distribution looks almost the same as Fig. \ref{fig:PC0_mcmc_r_w2} for all optical bands.

\begin{figure*}[p]
  \sbox0{\begin{tabular}{@{}c@{}}
    \includegraphics[width=\textheight]{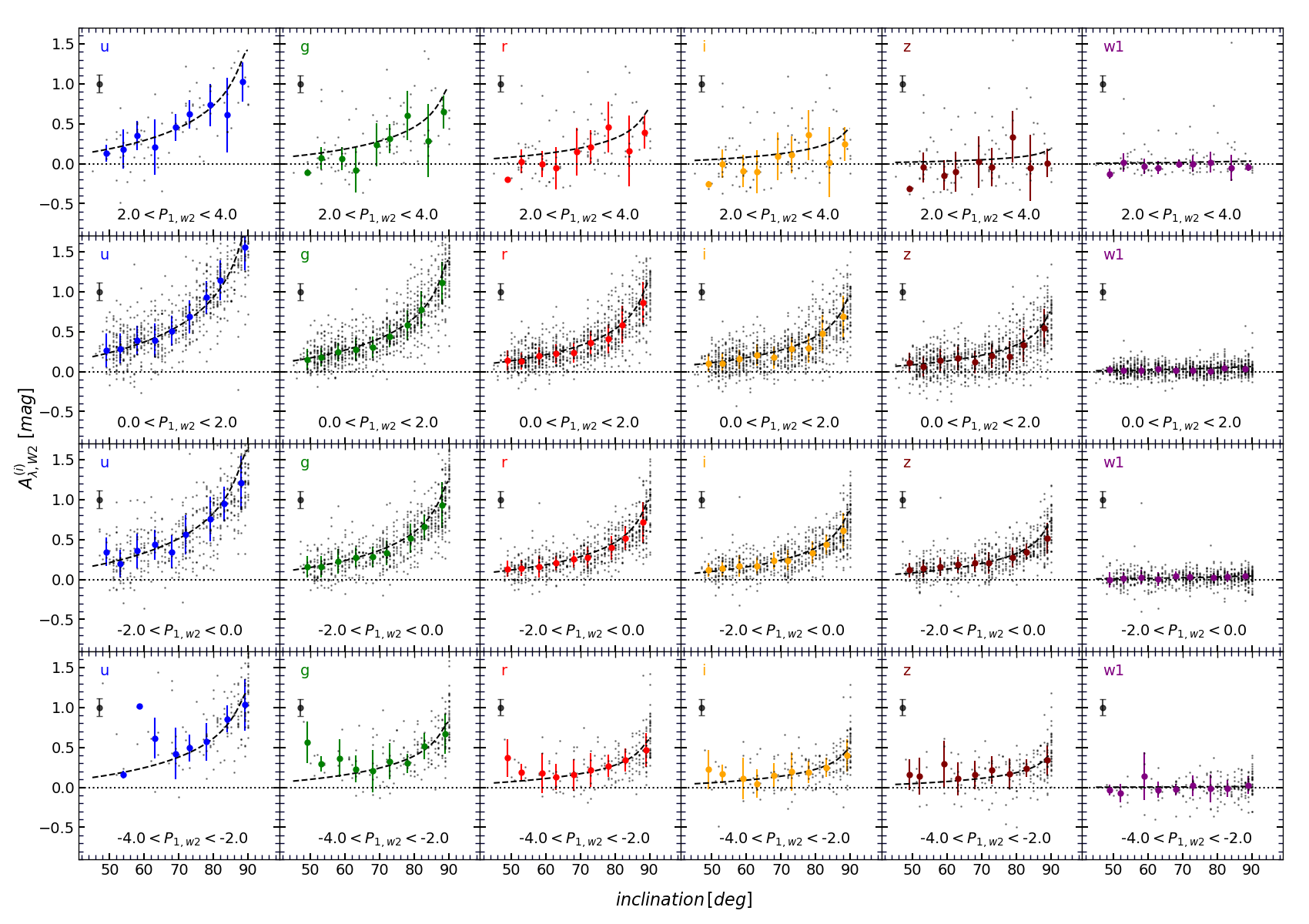}
  \end{tabular}}% measure width
  \rotatebox{90}{\begin{minipage}[c][\textwidth][c]{\wd0}
    \usebox0
\caption{Deviation of galaxies from the fiducial relation, Eq. \ref{Eq:GlWjO} due to their inclination, $A^{(i)}_{\lambda, W2}$, versus their inclination for different intervals of $P_{1,W2}$. Each gray dot represents an individual galaxy and color points are the average of data points in 5\dg inclination bins, with the error bars corresponding to the 1$\sigma$ scatter of data points. The black error barred black point on the top left corner of each panel display the median 1$\sigma$ uncertainty on the measured $A^{(i)}_{\lambda, W2}$ for the displayed spirals. Black dashed curves show our parametric model for the average main principal component of contributing galaxies in each panel. }
  \end{minipage}}
\label{fig:A_w2_inc}
\end{figure*}

\begin{figure*}[p]
  \sbox0{\begin{tabular}{@{}c@{}}
    \includegraphics[width=\textheight]{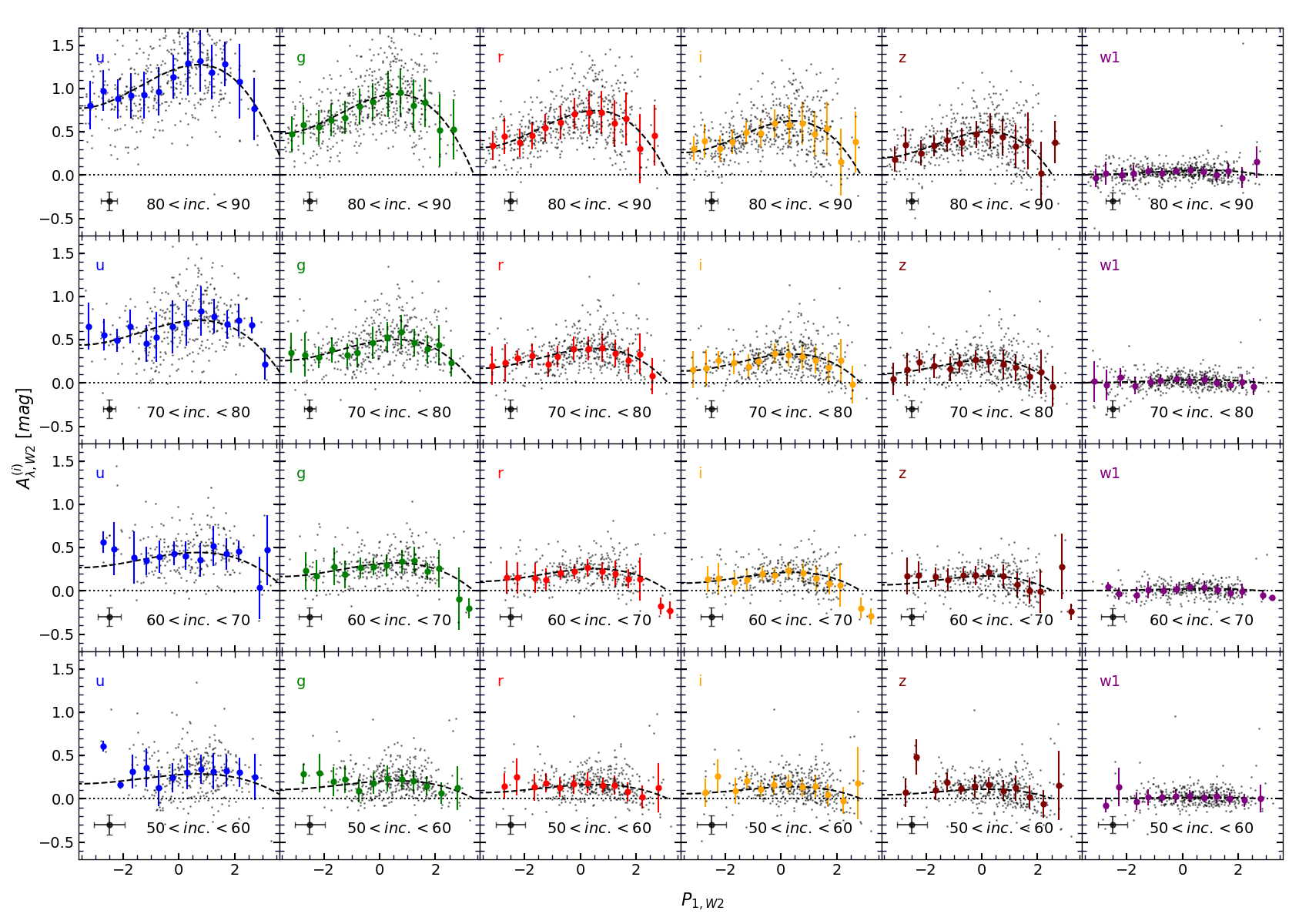}
  \end{tabular}}% measure width
  \rotatebox{90}{\begin{minipage}[c][\textwidth][c]{\wd0}
    \usebox0
    \caption{The reddening parameter $A_{\lambda, W2}^{(i)}$ vs. principal component parameter $P_{1,W2}$ for different inclination ranges. Color points represent the average of data points in horizontal bins with the size of $0.5$.}
  \end{minipage}}
\label{fig:A_w2_P0}
\end{figure*}

Fig. \ref{fig:color_pc0_w2_50-60} illustrates how spirals in the inclinations range between 50\dg and 60\dg are distributed relative to our model outcome for the linear fiducial relation between face-on color, $\Delta m_{\lambda J}^{(o)}$, and main principal component, $P_{1,W2}$, as defined by Eq. \ref{Eq:mWj}. 
Despite the similarity, we draw the reader attention to the difference between 
the presented fiducial lines in this diagram and that presented in \S \ref{sec:fcs}.

In Fig. \ref{fig:r_w2_features_Fon}, we only used face-on galaxies with no measurement for their inclinations, where we set all inclinations to 40\dg to explore the color dependency of face-on spirals on different parameters, by fitting the fiducial line using the plotted scattered points. However, in Fig. \ref{fig:color_pc0_w2_50-60}, we take the optimized parameters of our model (Table \ref{table:model_params}), which are found based on all galaxies with inclinations $>45$\dg to over-plot the fiducial relation on the scattered data points of spiral with inclinations that range between 50\dg and 60\dg.

\begin{figure*}[t]
\centering
\includegraphics[width=1.0\linewidth]{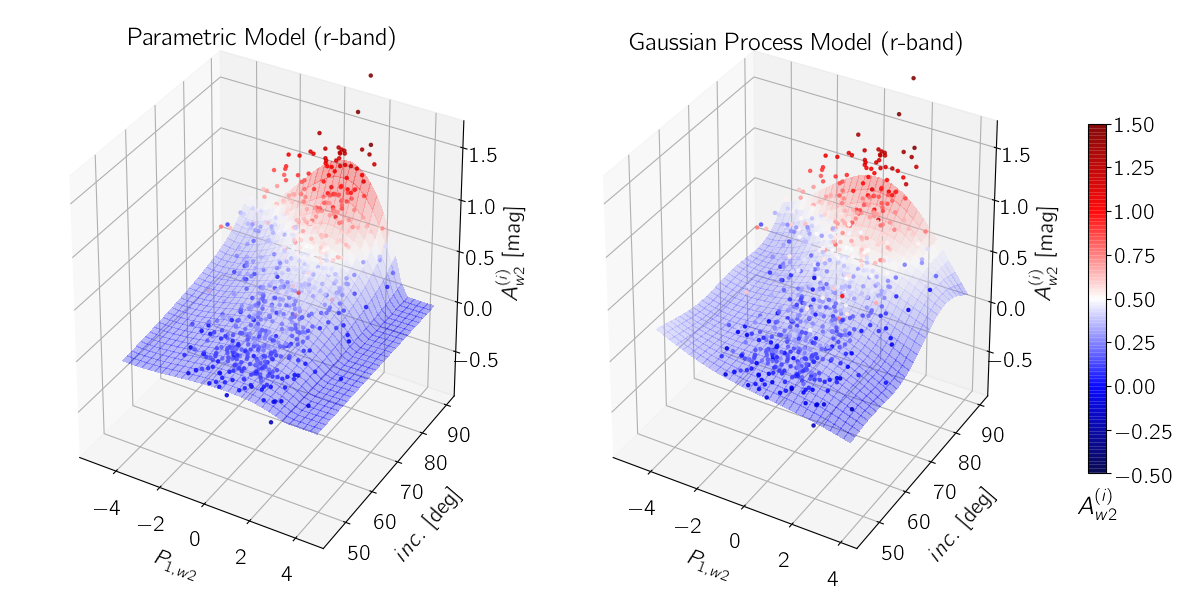}
\caption{The r-band inclination dependent dust obscuration in spiral galaxies, $A^{(i)}_{r,W2}$, as a function of their spatial inclination relative to the observer line-of-sight and their main principal component, $P_1$ (Eq \ref{eq:pca_combo}). {\bf Left:} The best fitted parametric model formulated by Eq. \ref{Eq:AWj} and parameters in Table \ref{table:model_params}. {\bf Right:} The non-parametric model based on a Gaussian Process formalism described in \S \ref{sec:gpc}. Each meshed surface shows the corresponding model whereas the colored points illustrate the location of galaxies in this space. To avoid confusion, only 700 data points are chosen randomly and over-plotted.}
\label{fig:Param_GP_rw2}
\end{figure*}

Fig. \ref{fig:A_w2_inc} shows different cross sections of our resulting models for $A^{(i)}_{\lambda, W2}$ at different wavelengths versus inclination for different ranges of $P_{1,W2}$. In these plots each gray dot represents a galaxy whose $A^{(i)}_{\lambda, W2}$ value is calculated following
\begin{equation}
\label{Eq:galaxy_individual}
A^{(i)}_{\lambda, W2} = (\overline{m}_\lambda - \overline{W}2)-(\alpha_\lambda P_{1,W2}+\beta_\lambda) , 
\end{equation}
where $\alpha_\lambda$ and $\beta_\lambda$ are the model parameters for the linear fiducial relation of face-on colors taken from Table \ref{table:model_params}. While $\alpha_\lambda$ and $\beta_\lambda$ are not purely observable quantities, they could be separately obtained from an independent analysis of fiducial linear relation of face-on spirals. Our roughly estimated $\alpha_\lambda$ and $\beta_\lambda$ values for close to face-on galaxies in \S \ref{sec:fcs} and what we get from our precise optimization of our parametric model (Table \ref{table:model_params}) point us to almost the same fiducial trends. As defined in Eq. \ref{Eq:galaxy_individual}, $A^{(i)}_{\lambda, W2}$ for each galaxy represents its color deviation from the fiducial relation of face-on spirals attributable to the inclination dependent dust attenuation.

In Fig. \ref{fig:A_w2_inc}, to better see the general trend of the data points, the color points show the mean of $A^{(i)}_{\lambda, W2}$ for galaxies in the inclinations bins of size 5\dg where the vertical error bars display 1$\sigma$ standard deviation of the scatter along the attenuation parameter. In each panel, the black dashed curve is the result of our model as we use the median of $P_{1,W2}$ value for all galaxies in that panel. The parameter $\mathcal{F}_\lambda(i)$ with fixed $q_{\lambda}$ controls how attenuation varies with inclination (see Eq. \ref{Eq:Fli}). As expected, the effect of dust attenuation decreases toward longer wavelengths, with minimal effect at $W1$-band. 

Similarly, Fig. \ref{fig:A_w2_P0} shows the behaviour of our model with variance of the principal component against data points in different intervals of inclinations. For a given inclination, $A^{(i)}_{\lambda, W2}$ is related to the main principal component through a polynomial function defined by Eq. \ref{Eq:GlWjI}, with a peak around $P_{1,W2}\simeq 1$. As expected, for a constant $P_{1,W2}$ value, the dust attenuation increases as galaxies become more edge-on.

The 3-dimensional behaviour of our model is seen in Fig. \ref{fig:Param_GP_rw2} where we plot $A^{(i)}_{r,W_2}$ as a function of inclination and the main principal component constructed from the $W2$-band photometry.  We set the attenuation to zero where the parametric model goes negative at $P_{1,W2}\gtrsim3$.

\subsection{Comparisons with a semi-parametric model} \label{sec:compare}

In \S \ref{sec:parametrization}, we constructed a parametric model of the effect of inclination on the fiducial color relations of the face-on galaxies. In this section, we take the Gaussian Process (hereafter GP) approach to predict $A^{(i)}_{\lambda J}$ for a set of parameters $P_{1,J}$ and inclination, based on the information provided by the observed data for neighbouring sample spirals in the parameter space. GP is a statistical machine learning technique based on a Bayesian approach that generates predictions following the behaviour of data \citep[e.g.][]{rasmussen:williams:2006, 2012MNRAS.419.2683G, 2017A&A...598A.125R}. Considering the strong correlations that underlie the fiducial relations for face-on spirals, we continue using Eq. \ref{Eq:GlWjO} with the best fitted $\alpha_{\lambda J}$ and $\beta_{\lambda J}$ taken from Table \ref{table:model_params} followed by a GP formalism to model $A^{(i)}_{\lambda J}$. See appendix \S \ref{sec:gpc} for the details of our GP algorithm setup.

Figure \ref{fig:Param_GP_rw2} compares the results of the parametric and GP models. Both models tend to underestimated a flare in the dust attenuation at $\sim$90\dg, although our parametric model appears to perform better at larger inclinations. In regions with numerous data points, say $-2<P_1<2$, the results of our alternate models are almost the same, however they differ significantly in less populated regions. The behavior of the parametric model is controlled by the adopted formulation that reasonably captures the physics of the problem, while the GP formalism solely learns from the distribution of data points to make predictions. Thus, one of the main downsides of the GP model in following the trend of data points is that it performs poorly wherever few measurements are available. The differences between parametric and GP models are modest but we give preference to the parametric model due to its simplicity.

\section{Dependency of dust attenuation on wavelength} \label{sec:rvw}

\begin{figure}[t]
\centering
\includegraphics[width=0.9\linewidth]{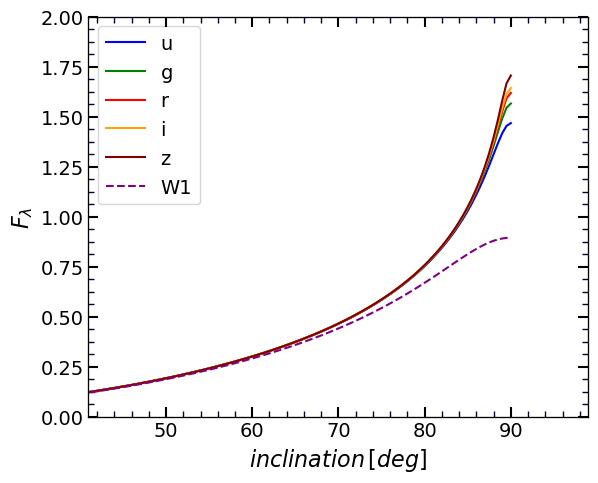}
\caption{The inclination dependent component of dust attenuation, $\mathcal{F_\lambda}(i)$, as defined in Eq. \ref{Eq:Fli}, with separate $q_\lambda$ values taken from Table \ref{table:model_params}.}
\label{fig:F_lambda_i}
\end{figure}

\begin{figure*}[ht]
\centering
\includegraphics[width=1.0\linewidth]{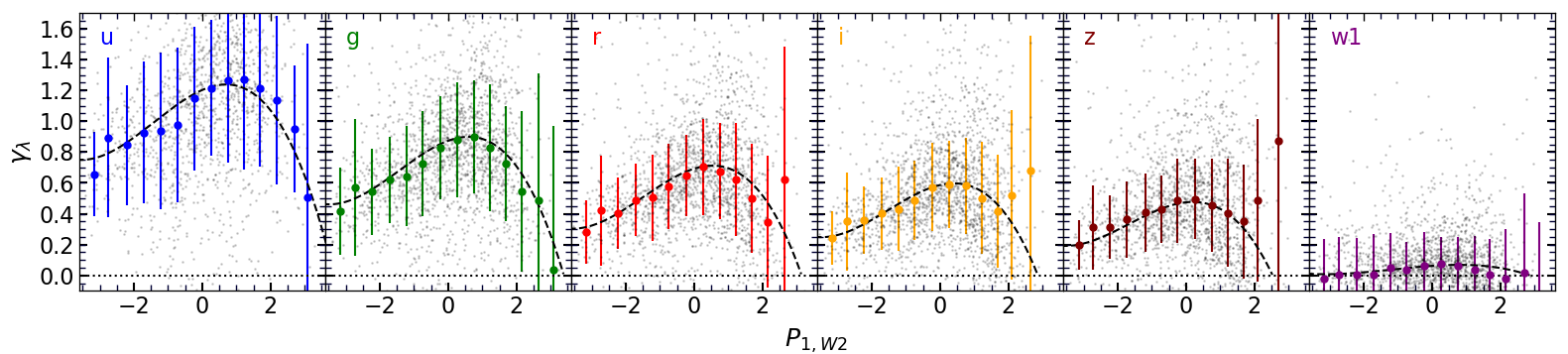}
\caption{Similar to Fig. \ref{fig:A_w2_P0} but for dust attenuation factor, $\gamma_{\lambda,W2}$. 
For each galaxy, $\gamma_\lambda$ is calculated from equations \ref{Eq:AWj}, \ref{Eq:Fli} and \ref{Eq:galaxy_individual}. Dashed curves display Eq. \ref{Eq:GlWjI} with the optimized parameters taken from Table \ref{table:model_params}.}
\label{fig:gamma_P1_lambda}
\end{figure*}

\begin{figure}[ht]
\centering
\includegraphics[width=0.80\linewidth]{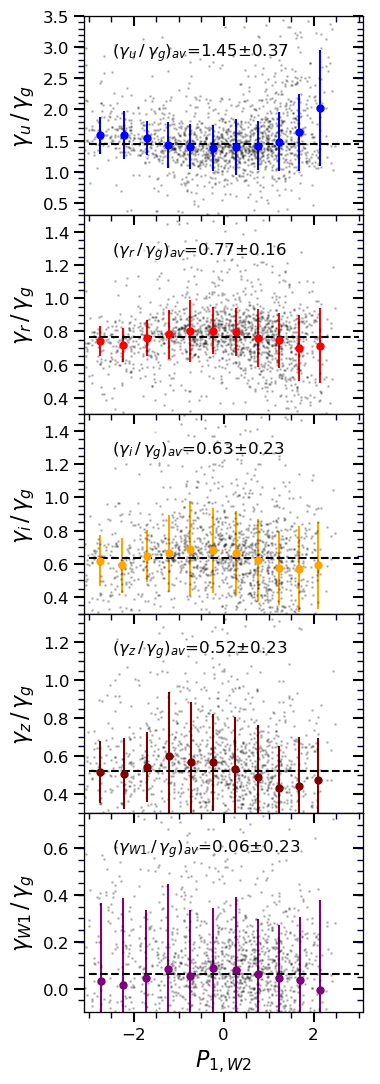}
\caption{Similar to Fig. \ref{fig:gamma_P1_lambda} but for ratio of dust attenuation in different wavebands. In each panel, the dashed horizontal line shows the average extinction ratio, $(\gamma_\lambda / \gamma_g)_{av}$, of all galaxies represented by gray dots.}
\label{fig:gamma_ratio}
\end{figure}

The inclination dependent dust attenuation was modelled independently in multiple passbands in \S \ref{sec:parametrization}.  The model follows Eq. \ref{Eq:AWj} and the best fitted parameters are presented in Table \ref{table:model_params}. In this equation, $\gamma_\lambda$ depends only on the observed properties of galaxies encoded in $P_{1,W2}$ and $\mathcal{F_\lambda}(i)$ is a function of galaxy inclination. Fig. \ref{fig:F_lambda_i} displays the behavior of $\mathcal{F_\lambda}(i)$ in different passbands. As seen, except for very edge-on spirals, at optical wavelengths $\mathcal{F_\lambda}$ is almost the same for all inclination values. We observe that the $W1$-band curve is shallower compared to those at optical bands, which is consistent with our expectation that dust absorption is less sensitive to inclination in longer wavelengths.

Absorption and inclination can be decoupled to give the physically meaningful dust attenuation factor $\gamma_{\lambda} = A_{\lambda, W2}^{(i)} /\mathcal{F}_\lambda(i)$.
In Fig. \ref{fig:gamma_P1_lambda}, we plot $\gamma_\lambda$ as a function of $P_{1,W2}$ in different wavebands. Each gray dot represents a spiral whose $\gamma_\lambda$ value is derived following equations \ref{Eq:AWj}, \ref{Eq:Fli} and \ref{Eq:galaxy_individual}. Dashed curves plot $\gamma_\lambda$ as parameterized in Eq. \ref{Eq:GlWjI} using the optimized parameters presented in Table \ref{table:model_params}. Consistent with our expectation, dust obscuration is greater in shorter wavebands with a similar dependency on $P_{1,W2}$ at all frequencies. We can see how dust attenuation in spiral galaxies varies with wavelength. Panels of Fig. \ref{fig:gamma_ratio} plot dust attenuation in spirals relative to that in $g$-band. In each panel, $\gamma_\lambda / \gamma_g$ for each galaxy is calculated by taking the corresponding $\gamma_\lambda$ values from Fig. \ref{fig:gamma_P1_lambda}. Color points and their error bars show median and 1$\sigma$ scatter of gray dots in bins with the size of 0.5 along $P_{1,W2}$. In each panel, a dashed horizontal line is drawn at the level of $(\gamma_\lambda / \gamma_g)_{av}$ which is derived by taking the median of all $\gamma_\lambda / \gamma_g$ fractions for all spirals. The scatter of color points and the size of their error bars do not imply any significant dependency of mean $\gamma_\lambda / \gamma_g$ value on $P_{1,W2}$.  

Dust extinction, $A_\lambda$, is usually normalized by color excess $E(B-V)\equiv A_B-A_V$
\begin{equation}
\kappa_\lambda = \frac{A_\lambda}{E(B-V)} = \frac{A_\lambda}{A_B-A_V} ~,
\end{equation}
where $\kappa_\lambda$ is the dust extinction factor. The value of $\kappa_\lambda$ in $V$-band for the Milky Way is $R_V \equiv \kappa_V=A_V/E(B-V)=3.1$ \citep{1989ApJ...345..245C}. According to this formalism, dust extinction ratios are given as
\begin{equation}
A_\lambda / A_V = \kappa_\lambda / \kappa_V = \kappa_\lambda / R_V.
\end{equation}

\begin{figure*}[ht]
\centering
\includegraphics[width=0.89\linewidth]{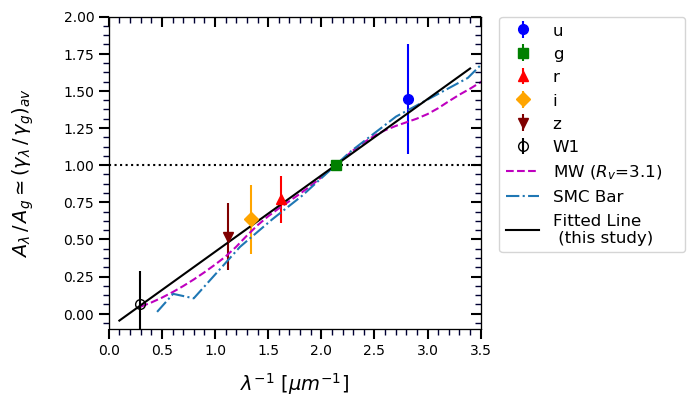}
\caption{Relative attenuation in different bands. attenuation values are normalized with respect to those in $g$-band. Magenta dashed curve shows the Milky Way dust extinction \citep{1989ApJ...345..245C}. Blue dashed dotted curve display the relative reddening in the Small Magellanic Cloud bar \citep{2003ApJ...594..279G}. Black solid straight line displays the best fitted linear relation given by Eq. \ref{Eq:gamma_ratio}.} 
\label{fig:gamma_ratio_lambda}
\end{figure*}

Adopting $(\gamma_\lambda / \gamma_g)_{av}$ values from Fig. \ref{fig:gamma_ratio}, we display the wavelength dependency of attenuation ratios, $A_\lambda / A_g \simeq (\gamma_\lambda / \gamma_g)_{av}$, in Fig. \ref{fig:gamma_ratio_lambda}.
As reference, reddening curves for the Milky Way \citep{1989ApJ...345..245C} and Small Magellanic Cloud (SMC) bar \citep{2003ApJ...594..279G} are over-plotted.

In Fig. 10 of their paper, \citet{2003ApJ...594..279G} present the reddening laws for several galaxies that suggests a linear dependency on the inverse of wavelengths in optical and infrared ranges. We find the following linear relationship for the average of our measured relative dust attenuation by fitting a straight line that passes through the $g$-band point
\begin{equation}
\frac{\gamma_\lambda}{\gamma_g} = (1.097\pm0.060)\Big[\frac{\lambda_g}{\lambda}-1\Big]+1 ~,
\label{Eq:gamma_ratio}
\end{equation}
where $\lambda_g = 4886 ~ \AA$ is the central wavelength of SDSS $g$-band filter. 
 Setting $\lambda_V=5500~\AA$, Eq. \ref{Eq:gamma_ratio} implies $\gamma_\lambda / \gamma_V = 1.194 (\gamma_\lambda / \gamma_g)$.

We note that at longer wavelengths our measured values for attenuation does not fall as rapidly as for the standard Galactic reddening case. This is in agreement with the average dust attenuation curve derived by \citet{2018ApJ...859...11S}, where a similar behavior is noticeable. The Galactic reddening curve pertains to the extinction over a range of wavelengths to a fixed source.  By contrast, the internal dust obscuration that we measure in galaxies averages over multiple paths to varying depths into a target that depend on wavelength.  At longer wavelengths galaxies become less optical thick and thus pass photons from more embedded regions \citep{1992ApJ...391..617H, 1998AJ....115.2264T}. Integrating the obscuration over a longer line-of-sight path consequently amplifies the effect of dust attenuation compared to Galactic extinction along a path to a fixed source. Furthermore, extinction involves photons scattered into the line of sight which can result in significant discrepancy between extinction and attention curves.

In appendix \ref{sec:surfB}, we find an empirical relation between $\gamma_\lambda$ and  effective surface brightness, $\langle \mu_2 \rangle^{(i)}_e$ which is useful when \hi 21 cm data is unavailable.

\section{Summary} \label{sec:summary}

Photons that are produced by the stellar content of a galaxy are absorbed and/or scattered by dust grains and gas molecules in its interstellar medium. Physical properties and chemical composition of dust particles in the interstellar medium of a galaxy depend on different factors like galaxy stellar ages, mass, morphology, metallicity, and formation history. The level of obscuration also depends on the spatial inclination of the galaxy relative to the observer. In galaxies with larger inclinations from face-on, the average photon travels along a longer line-of-sight path within the host and therefore has a higher chance to be obscured. 

The current study is an attempt to model the inclination dependent dust obscuration in spiral galaxies at multiple optical and infrared wavelengths. To acquire the necessary information to conduct this study, we performed photometry on a sample of 2,239 spirals with optical and infrared images provided by $SDSS$ and $WISE$ surveys. Another crucial piece of information is the inclinations of spirals which were carefully measured by visually comparing them to a set of standard spirals with known inclinations.

Empirically, larger galaxies have fractionally smaller HI components and, by inference, must be more efficient in consuming their \hi gas and forming stars. In the present universe, on average more massive galaxies tend to have older/redder stellar populations and have relatively less \hi gas. Prior to addressing the dependency of dust attenuation on inclination, we have established a set of inclination independent fiducial relations at different wavebands to capture this general color-size trend. We found relationships that describe the optical-infrared colors of face-on spirals in terms of other observable features that are correlated with their physical properties (\S \ref{sec:fcs} and \ref{sec:pca}). 

Turning to spirals with greater inclinations, we then model the inclination dependency of deviations of optical-infrared colors from the fiducial relations. These departures are attributed to the inclination dependent effects of dust obscuration on the observables.

In \S \ref{sec:parametrization} we established a parametric model to describe attenuation as a function of inclination and other observables. Taking a non-parametric approach in \S \ref{sec:gpc}, we used the Gaussian Process formalism to study dust obscurations, and we show the differences between our two methods are not significant.

Phenomenologically, both dwarf galaxies and galaxies proceeding toward red and dead have negligible optical attenuation.  On the other hand, intermediate spiral galaxies can loose well more than half their light to attenuation at blueward bands.  Low mass galaxies can have fractionally large amounts of \hi but this component of interstellar material is optically transparent.  Small galaxies have low metal content and have very limited molecular gas reservoirs \citep{2012MNRAS.422..215Y, 2016MNRAS.455.1156B}. 
Dust grains in the interstellar medium are the products of the release of metals due to star formation activity and the late stage evolution of stars. In the regime more massive than dwarfs that interests us, there is a roughly linear logarithmic relation between dust-to-gas ratio and metallicity \citep{2011ApJ...737...12L, 2015MNRAS.449.3274F}. Global galaxy metallicity can serve as a proxy of galaxy dust levels \citep{2019MNRAS.484.2587M}. Mean metallicity increases with galaxy mass until it saturates and grows only slowly for galaxies more massive than $M_* \simeq 10.5 M_\odot$ \citep{2004ApJ...613..898T,2014ApJ...791..130Z}. 

A generally consistent picture emerges.  Proceeding from smaller to larger masses, the metal content of the interstellar medium increases within abundant \hi reservoirs, resulting in increasing attenuation from opaque nebulosity.  At some point in this progression to more massive systems, though, the level of metallicity no longer grows significantly while the \hi content becomes increasingly depleted. Stellar winds or active galactic nuclei are increasing effective at clearing the galaxy of interstellar medium \citep{2011MNRAS.417.2962P,2015MNRAS.450..342K}. 

Our Principle Component parameter $P_{1,J}$ captures the essence of this interplay.  This parameter is a linear combination of three distance independent observables: (1) the amplitude of galactic rotation which is a proxy for total galaxy mass, (2) an \hi to infrared flux pseudo-color that monitors the relative amount of gas available to form young stars vs. the quantity of old stars, and (3) the infrared surface brightness.  Dust attenuation grows with $P_{1,J}$ until it reaches a maximum and turns over around $P_{1,J} \simeq 1$.  Increasing galaxy mass is a factor causing increasing obscuration while an increasingly red pseudocolor $C_{21W2}$ is a factor in diminishing obscuration.  Interestingly, the single parameter that comes closest to capturing the essence of the $P_{1,J}$ correlation with attenuation is the infrared surface brightness.

The differential reddening from SDSS $u$ band to WISE $W2$ band is only slightly different from the familiar Galactic reddening law with $R_V=3.1$.  As can be anticipated, mean reddening is slightly greater at longer wavelengths than anticipated by the Galactic reddening law to a fixed source because separate lines-of-sight penetrate hosts to different depths as a function of wavelength. 

\section*{Acknowledgments}
\acknowledgments
We are pleased to acknowledge the citizen participation to scientific research of undergraduate students at University of Hawaii, members of amateurs astronomy clubs in France Plan\'etarium de Vaulx-en-Velin, Association Clair d'\'etoiles et Brin d'jardin, Soci\'et\'e astronomique de Lyon, Club d'astronomie Lyon Amp\`ere, Club d'astronomie des monts du lyonnais, Club d'astronomie de Dijon, and friends who helped us with measuring inclinations of spiral galaxies in our sample.

Support for EK and RBT was provided by NASA through grant number 88NSSC18K0424
from the Space Telescope Science Institute. 
HC acknowledges support from Institut Universitaire de France.

This research has made use of the NASA/IPAC Extragalactic Database\footnote{\url{ttp://nedwww.ipac.caltech.edu/}} which is operated by the Jet Propulsion Laboratory, California Institute of Technology, under contract with the National Aeronautics and Space
Administration. This research made use of Montage, funded by the National Aeronautics and Space Administration’s Earth Science Technology Office, Computational Technologies Project, under Cooperative Agreement Number NCC5-626 between NASA and the California Institute of Technology. The code is maintained by the NASA/IPAC Infrared Science Archive.

\appendix

\section{Gaussian process recipe to model $A^{(i)}_{\lambda J}$} \label{sec:gpc}

Following the GP algorithm, we assume that $N$ observations for $A^{(i)}_{\lambda J}$ are drawn from a prior normal Gaussian distribution, i.e. $A^{(i)}_{\lambda J} \sim \mathcal{N}(0,\mathcal{K})$, where the mean is zero and $\mathcal{K}$ is the $N\times N$ covariance matrix generated using the uncertainty of $A^{(i)}_{\lambda J}$ and the GP kernel function that is chosen based on the expected behaviour of data points. Let {\bf A} be a $N\times 1$ column vector whose elements are $A^{(i)}_{\lambda J}$ measured for $N$ spirals. For any pair of galaxies, the covariance matrix element is written as 

\begin{equation}
\label{Eq:GP_covariance}
\mathcal{K}_{ab}=\sigma^2_{A_a} \delta_{ab}+\kappa_{ab}~, ~~ (a,b=1,2,...,N)
\end{equation}
where $\sigma_{A_a}$ is the uncertainty of $A_a$, $\delta$ is the Kronecker delta ($\delta_{ab}=1$ if $a=b$ and $\delta_{ab}=0$ if $a\neq b$) and $\kappa$ is the GP kernel. Considering the non-parametric nature of our problem, we adopt an exponential squared kernel function defined as 

\begin{equation}
\label{Eq:GP_kernel}
\kappa_{ab}=\sigma^2_e \delta_{ab}+\sigma^2_f~\exp\bigg[-\Big(\frac{P_{1a}-P_{1b}}{\ell_0}\Big)^2-\Big(\frac{i_a-i_b}{\ell_1}\Big)^2\bigg]  ~,
\end{equation}
where $P_1$ is the main principal component, $i$ is inclination, and $\sigma_e$, $\sigma_f$, $\ell_0$ and $\ell_1$ are the hyper-parameters of the chosen kernel. 
Training this GP model involves tuning the kernel hyper-parameters by maximizing the probability of {\bf A} to be drawn from the prior distribution, $\mathcal{N}(0,\mathcal{K})$. Similar to Eq. \ref{Eq:likelihood1}, the logarithm of likelihood has the following form  
\begin{equation}
\label{Eq:likelihood2}
\log \Lc({\bf A})= -\frac{1}{2}{\bf A}^T \mathcal{K}^{-1}{\bf A}-\frac{1}{2}\log |\mathcal{K}|-\frac{N}{2}\log (2 \pi) .
\end{equation}

Adopting the above likelihood and following the same approach explained in \S \ref{sec:opt}, we sample the posterior distribution of the hyper-parameters by performing MCMC simulations with 64 chains, each randomly initialized and sampled 10,000 times. We ignore the first 1,000 samples of each chain to ensure they are converged. Table \ref{table:hyperparams} lists the optimized hyper-parameters and their uncertainty bounds after merging all sampled chains.

\begin{table}[ht]
\setlength{\tabcolsep}{0.3cm}
\centering
\caption{Optimized hyper-parameters for the inverse squared GP kernel defined in Eq. \ref{Eq:GP_kernel} to model the inclination dependent dust attenuation in spirals, $A^{(i)}_{\lambda J}$ at optical wavelengths. 
\label{table:hyperparams}}
\begin{tabular}{ccccc}
\hline
$\lambda$ & $\log (\ell_0)$ & $\log (\ell_1)$ & $\log (\sigma^2_f)$ & $\sigma^2_e$ \\
\hline \hline
$u$ & $3.79^{+0.58}_{-0.50}$ & $6.94^{+1.09}_{-0.76}$ & $0.90^{+1.11}_{-0.85}$ & $0.165\pm0.005$  \\[1.5ex] 
$g$ & $3.26^{+0.49}_{-0.42}$ & $5.82^{+0.65}_{-0.58}$ & $-0.31^{+1.00}_{-0.77}$ & $0.110\pm0.004$  \\[1.5ex] 
$r$ & $3.08^{+0.47}_{-0.41}$ & $5.43^{+0.63}_{-0.55}$ & $-0.88^{+0.96}_{-0.74}$ & $0.097\pm0.003$  \\[1.5ex]
$i$ & $2.91^{+0.47}_{-0.41}$ & $5.17^{+0.65}_{-0.57}$ & $-1.39^{+0.94}_{-0.71}$ & $0.099\pm0.003$  \\[1.5ex]
$z$ & $2.86^{+0.48}_{-0.43}$ & $4.93^{+0.76}_{-0.63}$ & $-1.90^{+0.89}_{-0.68}$ & $0.099\pm0.003$ \\[1.5ex]
\hline
\end{tabular}
\end{table}

% \subsection{Gaussian Process predictions}
After finding the functional form of the GP model by optimizing its kernel hyper-parameters based on the data of $N$ galaxies with the parameters ${\bf A}$, ${\bf P_1}$ and ${\bf i}$, we need a formalism to predict ${\bf A_*}$ for a set of $M$ galaxies with measured ${\bf P_{1*}}$ and ${\bf i_*}$. For simplicity we define ${\bf X}$ and ${\bf X_*}$ to be the input parameters defined as $({\bf P_1}$,${\bf i})$ and $({\bf P_{1*}}$,${\bf i_*})$ respectively. Based on the GP assumption, the joint probability of ${\bf A}$ and ${\bf A_*}$ should follow the same Gaussian distribution, therefore ${\bf A}, {\bf A_*}|{\bf X},{\bf X_*} \sim \mathcal{N}(0,\mathcal{K_+})$, where $\mathcal{K_+}$ could be expressed as

\begin{equation}
\label{Eq:Kplus}
\mathcal{K_+}= \begin{pmatrix} \mathcal{K}({\bf X},{\bf X}) &  \mathcal{K}({\bf X},{\bf X_*})\\  \mathcal{K}({\bf X_*},{\bf X}) & \mathcal{K}({\bf X_*},{\bf X_*}) \end{pmatrix},
\end{equation}
where $\mathcal{K}({\bf X},{\bf X})$ is $N\times N$, $\mathcal{K}({\bf X_*},{\bf X_*})$ is $M\times M$, $\mathcal{K}({\bf X},{\bf X_*})$ is $N\times M$ and equals $\Big(\mathcal{K}({\bf X_*},{\bf X})\Big)^T$.  ${\bf A_*}|{\bf A},{\bf X},{\bf X_*}$ also follows a Gaussian distribution $\mathcal{N}(\mu^*,\Sigma^*)$ where its mean and covariance matrix are given by

\begin{equation}
\label{Eq:GPposterior}
\begin{split} 
\mu^* &= \mathcal{K}({\bf X_*},{\bf X}) \Big(\mathcal{K}({\bf X},{\bf X})\Big)^{-1}{\bf A} ~, \\
\Sigma^* &= \mathcal{K}({\bf X_*},{\bf X_*})
   -\mathcal{K}({\bf X_*},{\bf X})\Big(\mathcal{K}({\bf X},{\bf X})\Big)^{-1}\mathcal{K}({\bf X},{\bf X_*}).
\end{split}
\end{equation}

We use the George Python package developed by \citet{2015ITPAM..38..252A} to facilitate the calculations of the kernel matrix defined by Eq. \ref{Eq:GP_kernel} and to make predictions based on Eq. \ref{Eq:GPposterior}. The right panel of Fig. \ref{fig:Param_GP_rw2} displays the prediction of our resulting GP model for $A^{(i)}_{r,W_2}$. 

\section{Dust attenuation as a function of infrared surface brightness} \label{sec:surfB}

In \S \ref{sec:fcs} and \ref{sec:pca}, we study the correlations between various distance independent features and optical-infrared color, $\overline{m}_\lambda - \overline{W}j$. Throughout the paper, we conducted our analysis of dust attenuation using the main principal component, $P_1$, based on its strongest correlation with optical-infrared colors. We used $P_1$ as a proxy that effectively includes the influential physical property of spirals of importance in our study. Calculation of $P_1$ requires having knowledge of the \hi line profile and its flux. Our sample spirals are chosen to have observations at 21 cm wavelength for the purpose of distance measurements. In our study, \hi information is an essential ingredient in the calculation of dust attenuation. However, the 21 cm profile/flux might not be available for an arbitrarily chosen spiral. Here, we attempt to present an empirical function for dust attenuation that solely relies on optical and infrared photometric information.

\begin{figure}[t]
\centering
\includegraphics[width=\linewidth]{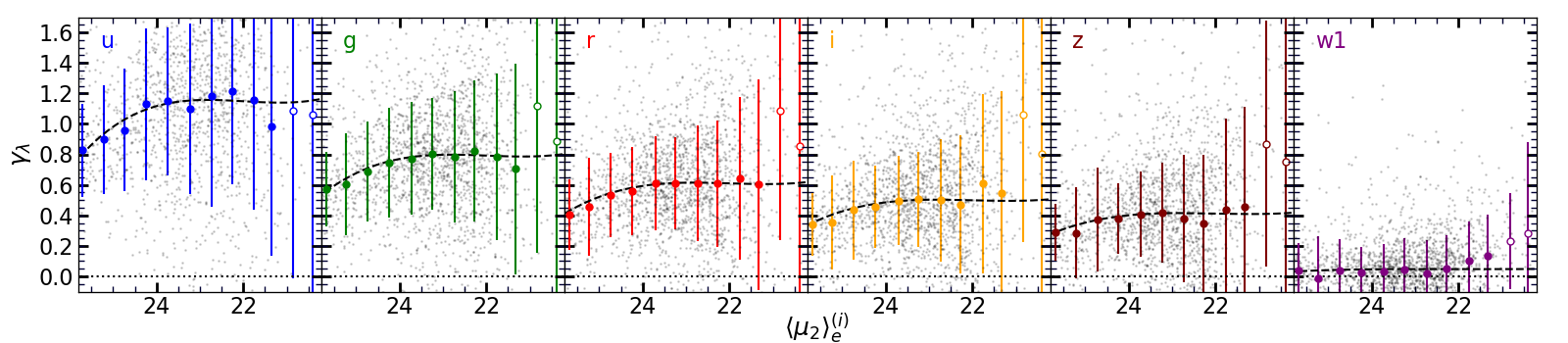}
\caption{Similar to Fig. \ref{fig:gamma_P1_lambda}, dust extinction, $\gamma_{\lambda,W2}$ vs. the $W2$ effective surface brightness, $\langle \mu_2 \rangle^{(i)}_e$. Each gray dot represents a spiral whose $\gamma_\lambda$ from equations \ref{Eq:mWj}-\ref{Eq:Fli}. The color points display the average location of galaxies within surface brightness bins with the size of $0.5$ mag, and open symbols represent surface brightness values brighter than $21$ mag. Black curves are the best fitted third degree polynomials to gray points that is given by Eq. \ref{Eq:gamma_surfB}.
}
\label{fig:surfB_w12}
\end{figure}

The strong correlation between effective surface brightness, $\langle\mu_2\rangle^{(i)}_e$, $\overline{r}-\overline{W}2$ and $C_{21W2}$ (Fig. \ref{fig:corr_table}) motivates us to search for a relationship between dust attenuation and infrared surface brightness. Fig. \ref{fig:surfB_w12} display $\gamma_{\lambda}$ vs. $\langle\mu_2\rangle^{(i)}_e$ at different wavebands. Given $P_1$ for each spiral, $\gamma_\lambda$ is calculated following Eq. \ref{Eq:GlWjI} with parameters taken from Table \ref{table:model_params}. Figures \ref{fig:gamma_P1_lambda} and \ref{fig:surfB_w12} motivate us that the average dependency of $\gamma_\lambda$ on the physical observables of spirals could possibly be described using similar functions in different wavebands. With that in mind and for the sake simplicity, we adopt this relation for the dust extinction: $\gamma_{\lambda}\triangleq\rho_\lambda G$, where $G$ is a function of infrared surface brightness and has the same form at all bands, and $\rho_\lambda$ is the wavelength dependent dust attenuation factor. 

We use a third degree polynomial function to describe $G$. To obtain more robust results and to improve the statistics, our fitting procedure includes all data points in multiple $u,g,r,i,z$ and $W1$ bands simultaneously. Our optimized relation between $\gamma_\lambda$ and the effective surface brightness of spirals in $W2$-band is given as

\begin{equation}
\gamma_{\lambda,W2} =\rho_\lambda \Big( -5.4407\times 10^{-3}\langle \mu_2 \rangle^3_e+0.35885\langle \mu_2 \rangle^2_e-7.8782\langle \mu_2 \rangle^2_e+58.363 \Big) ~,
\label{Eq:gamma_surfB}
\end{equation}
where $\langle \mu_2 \rangle_e$ is the inclination corrected effective surface brightness and $\rho_\lambda \simeq (\gamma_\lambda / \gamma_g)_{av}$ is the dust extinction factor normalized to $g$-band (see \S \ref{sec:rvw}). The values of $\rho_\lambda$ for $u,r,i,z$ and $W1$ bands are $1.45\pm0.37$, $0.77\pm0.16$, $0.63\pm0.23$, $0.52\pm0.23$ and $0.06\pm0.23$ respectively (see Fig. \ref{fig:gamma_ratio}).

\section{$W1$ vs. $W2$ principal components} \label{sec:w1/w2}

\begin{figure*}[t]
\includegraphics[width=\linewidth]{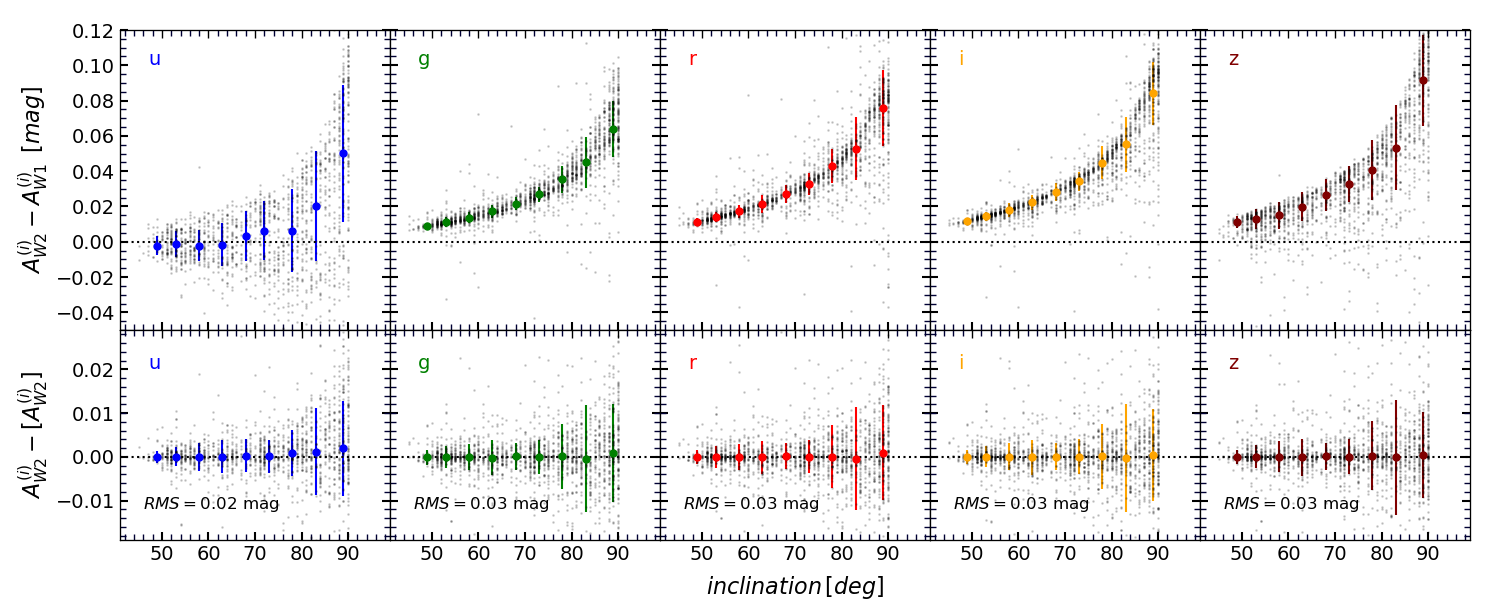}
\caption{{\bf Top row:} The discrepancy between the inclination dependent attenuation in spirals calculated based on $W1$ and $W2$-band main principal components, $A^{(i)}_{W2}-A^{(i)}_{W1}$, versus inclination in different optical bands. {\bf Bottom row:} The same as the top row but for the difference between $A^{(i)}_{W2}$ and the estimated dust attenuation, $[A^{(i)}_{W2}]$, calculated using the linear relation presented in Fig. \ref{fig:P0_w12} to derive $P_{1,W2}$ from $P_{1,W1}$. Each gray dot represents a galaxy and the points with error bars show the average location of galaxies within inclination bins with the size of 5\dg where error bars show 1$\sigma$ scatter of vertical position of galaxies within each bin.}
\label{fig:A_w12_inc}
\end{figure*}

\begin{figure}[t]
\centering
\includegraphics[width=0.55\linewidth]{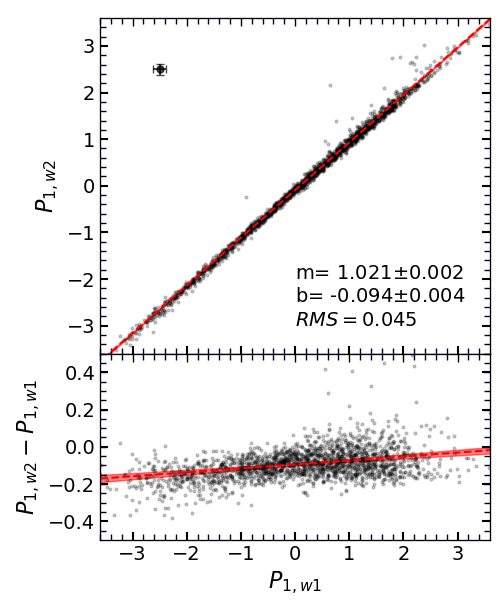}
\caption{First principal component calculated using $W2$-band photometric data, $P_{1,W2}$ versus that obtained from the $W1$-band images, $P_{1,W1}$. Each black dot represent a galaxy with inclination $>45$\dg. The red line shows the best fitted line with the slope {\it m} and intercept {\it b}. The RMS of deviations of black points from the fitted line is calculated along the vertical axis, $P_{1,W2}$}
\label{fig:P0_w12}
\end{figure}

One of the most important parameters in our analysis is the main principal component that can be derived based on the photometry of either $W1$ or $W2$ bands. 
Up to this point in this paper, dust obscuration in optical bands and $W1$-band are derived using the $P_{1,W2}$ parameter which is believed to not be significantly under the influence of dust attenuation. Although the effect of dust obscuration is very small on $W1$-band fluxes, it is not negligible. Despite the small slope of the fiducial relation for the $\overline{W}1-\overline{W}2$ color (Fig. \ref{fig:color_pc0_w2_50-60}) that suggests the possibility of obtaining the same estimation for the inclination depend attenuation following the same analysis using $P_{1,W1}$ instead of $P_{1,W2}$, the outcome might be biased. In the top row of Fig. \ref{fig:A_w12_inc}, we plot the difference between the outputs of our parametric model for dust attenuation that are derived based on the $W1$ and $W2$-band luminosities, i.e. $A^{(i)}_{\lambda, W1}$ and $A^{(i)}_{\lambda, W2}$, versus inclination. As expected, this deviation becomes more pronounced as inclination increases, because $W1$-band luminosities are still prone to dust obscuration in more edge-on spirals. Looking at the $W1$ panels of Fig. \ref{fig:A_w2_inc}, one might notice this minor inclination dependency of the deviation of $\overline{W}1-\overline{W}2$ colors from the corresponding fiducial relation.

On the other hand, the higher quality of photometric data at $W1$-band motivates us to build a framework based on $W1$ band photometry which does not suffer such an inclination dependent bias. Thus, it is more reasonable to use the same formalism we obtained for the $W2$ band that uses $P_{1,W2}$. 

Fig. \ref{fig:P0_w12} plots $P_{1,W1}$ versus $P_{1,W2}$ for all sample spirals suggesting a linear relation that allows us to estimate the $W2$ principal component based on $W1$ band photometric data.
The best fitted relation is $[P_{1,W2}]=1.021 P_{1,W1}-0.094$, where $[P_{1,W2}]$ represents the approximated value of $P_{1,W2}$. We feed the estimated $[P_{1,W2}]$ to the formalism generated based on $W2$-band to determine the dust attenuation, $[A^{(i)}_{W2}]$. In the bottom row of Fig. \ref{fig:A_w12_inc}, we plot the difference between this new estimation of attenuation, which relies on the $W1$-band, and $A^{(i)}_{W2}$ versus inclination. For all wavelengths, the RMS scatter of deviations do not show any significant inclination dependant variations. Over all inclinations the scatter of deviations is less than $\sim 0.03$ {\rm mag} with the larger scatter for more edge-on galaxies.

\bibliography{paper}

\end{document}